\documentclass{elsart5p}
\usepackage{graphicx,amssymb,lineno}
\def\bm#1{\mathchoice
 {\mbox{\boldmath$\displaystyle#1$}}%
 {\mbox{\boldmath$#1$}}%
 {\mbox{\boldmath$\scriptstyle#1$}}%
 {\mbox{\boldmath$\scriptscriptstyle#1$}}}
\newcommand{\ud}{\mathrm{d}}
\newcommand{\ue}{\mathrm{e}}
\newcommand{\tr}{\mathrm{tr}\,}

\begin{document}

\begin{frontmatter}

\title{Recent~Fluid~Deformation~closure for~velocity~gradient~tensor~dynamics in~turbulence: time-scale~effects~and~expansions}
\author{Marco~Martins~Afonso\thanksref{tha1}}
 \thanks[tha1]{Supported by Keck Foundation.}
  \&
\author{Charles~Meneveau\thanksref{tha2}\corauthref{cor1}}
 \ead{meneveau@jhu.edu}
 \thanks[tha2]{Supported by NSF (AST-0428325).}
 \corauth[cor1]{Corresponding author.}
 \address{Department of Mechanical Engineering, Johns Hopkins University, 3400 North Charles Street, Baltimore, MD 21218 (USA)}

 \begin{abstract}
  In order to model pressure and viscous terms in the equation for the Lagrangian dynamics of the velocity
  gradient tensor in turbulent flows, Chevillard \& Meneveau ({\it Phys.\ Rev.\ Lett.}\ {\bf 97}, 174501, 2006)
  introduced the Recent Fluid Deformation closure.
  Using matrix exponentials, the closure allows to overcome the unphysical finite-time blow-up of the well-known Restricted Euler model.
  However, it also requires the specification of a decorrelation time scale of the velocity gradient along the Lagrangian evolution,
  and when the latter is chosen too short (or, equivalently, the Reynolds number is too high),
  the model leads to unphysical statistics. In the present paper, we explore the limitations of this closure
  by means of numerical experiments and analytical considerations. We also study the possible effects of using
  time-correlated stochastic forcing instead of the previously employed white-noise forcing.
  Numerical experiments show that reducing the correlation time scale specified in the closure and in the forcing does not
  lead to a commensurate reduction of the autocorrelation time scale of the predicted evolution of the velocity gradient tensor.
  This observed inconsistency could explain the unrealistic predictions at increasing Reynolds numbers.
  We perform a series expansion of the matrix exponentials in powers of the decorrelation time scale,
  and we compare the full original model with a linearized version.
  The latter is not able to extend the limits of applicability of the former but allows the model to be cast
  in terms of a damping term whose sign gives additional information about the stability of the model as function of the
  second invariant of the velocity gradient tensor.
 \end{abstract}


\end{frontmatter}

 \section{Introduction}

  The study of the velocity gradient tensor $A_{ij}\equiv\partial u_i/\partial x_j$, where
  $\bm{u}(\bm{x},t)$ is the velocity field, has great significance in determining the small-scale
  structure of turbulent flows. Among others, it allows to identify the areas of the flow in which either
  strain-rate (i.e.\ deformation) or vorticity (i.e.\ rotation) prevails. It contains geometric information about the orientation
  of vorticity and strain-rate eigenvectors, and its higher-order moments quantify the level of intermittency of
  small-scale turbulence. Here we focus on three-dimensional, incompressible
  turbulent flows, i.e.\ $\bm{\partial}\cdot\bm{u}=A_{ii}=0$.

  The Lagrangian interpretation of the evolution equation for $\bm{A}$,
  derived from the (Eulerian) Navier--Stokes equation,
  leads to a closure problem due to the fact that the pressure Hessian and viscous term are not directly
  expressed in terms of the local value of $\bm{A}$.
  The Restricted Euler (RE) model \cite{V84,C92} is the classical
  and crudest closure scheme, as it completely neglects the viscous
  term and the deviatoric part of the pressure Hessian, assuming
  the latter to be isotropic. The RE closure captures several qualitative features
  of real turbulence, such as the alignment of the vorticity vector
  with the direction corresponding to the intermediate eigenvalue of
  the strain matrix, and the statistical prevalence of axisymmetric
  expansion (situations in which a fluid element, initially spherical,
  become pancake-like rather than cigar-like).
  It allows the analytical investigation of the Lagrangian evolution
  of material volume elements and of their deformation tensor \cite{LM07}.
  Moreover, it also represents the starting point to deduce
  the so-called Advected Delta-Vee (ADV) system for velocity increments
  \cite{LM05,LM06}. However, it presents a major drawback,
  namely the appearance of an unphysical finite-time singularity.

  With the aim of overcoming this problem, the so-called Recent Fluid Deformation
  (RFD) closure was introduced \cite{CM06,CM07,CMBT08}. It provides a model
  for the pressure Hessian and viscous term, which are suitably parameterized,
  i.e.\ expressed as functions of the local velocity gradient tensor. Using matrix exponentials, it
  takes into account both the geometry and the dynamics of the recent history
  of the deformation of a fluid element, and requires the specification of a
  decorrelation time scale $\tau$, of the order of the Kolmogorov time scale.
  It was shown that this refined parameterization leads to stationary statistics,
  removing the above-mentioned finite-time singularity. Moreover, it also captures some other
  relevant properties observed in real turbulence, such as: the non-Gaussianity
  of longitudinal and transverse velocity gradients, the characteristic teardrop
  shape of the probability density function (PDF) in the  $R$-$Q$ plane,
  the quasi-lognormality of pseudodissipation, and, more importantly for our purposes,
  the correct scaling with the Reynolds number ($\mathrm{Re}\sim\tau^{-2}$) at small to
  intermediate values of $\mathrm{Re}$. However, when $\tau$ is chosen too short (or,
  equivalently, $\mathrm{Re}$ is too high), unphysical statistics are observed in the model \cite{CM06,CM07}.

  In the present paper, we investigate the limitations of the RFD closure at increasing
  Reynolds numbers (or decreasing imposed correlation time scale, $\tau$)
  by means of numerical computations. In particular, the predicted Lagrangian
  time-correlation structure of the model is studied. We also propose a
  simplified, linearized version of the model, which turns out to have the same (or even narrower)
  ranges of applicability,  but allows us to proceed analytically (in the spirit of
  \cite{MDV98}) and to draw some additional conclusions.

  The paper is organized as follows. In section \ref{sec:RERFD}
  we review the RE model and RFD closure.
  The equation deduced from RFD is then integrated numerically in time,
  and the results are shown in section \ref{sec:num}.
  In section \ref{sec:exp} we perform an analytical manipulation
  of the matrix exponential, expanding it in a power series in $\tau$,
  and we show the resulting equations for $\bm{A}$ at the different orders in $\tau$.
  In section \ref{sec:lin} we focus on the $\tau$-linear
  approximation and we deduce a restricted dynamical system.
  Conclusions and perspectives follow in section \ref{sec:conc}.
  Appendix \ref{sec:app} shows some details about the random forcing
  term we use to obtain stationary statistics for the RFD closure.

 \section{Review of RE model and RFD closure} \label{sec:RERFD}

  Starting from the Navier--Stokes equation,
  \[\partial_tu_i+u_k\partial_ku_i=-\partial_ip+\nu\partial^2u_i\]
  ($p$ is the pressure divided by density and $\nu$ the kinematic viscosity),
  it is sufficient to take the gradient ($\partial_j$)
  to obtain the evolution equation for the velocity gradient tensor:
  \begin{equation} \label{evol}
   \partial_tA_{ij}+u_k\partial_kA_{ij}=-A_{ik}A_{kj}-\partial_i\partial_jp+\nu\partial^2A_{ij}\;.
  \end{equation}
  The incompressibility constraint makes $\bm{A}$ a traceless tensor,
  $\tr\bm{A}=0$, thus reducing its number of independent components from 9 to 8.

  Equation (\ref{evol}), interpreted in Eulerian sense, tells us little
  more than the original Navier--Stokes formulation. The key point consists in
  reinterpreting it in a Lagrangian sense, by considering its left-hand side
  as a material derivative following a fluid particle. However, one immediately
  faces a particular closure problem, because the last two term at its right-hand
  side (RHS) are not known in terms of $\bm{A}$ at the same point and time.
  The isotropic part of the pressure Hessian can
  easily be rewritten by using the Poisson equation (obtained from (\ref{evol}) by
  taking its trace), which allows to express the pressure Laplacian as a
  quadratic term in $\bm{A}$: $\partial^2p=-\tr\bm{A}^2$. Thus, one obtains:
  \begin{equation} \label{vgt}
   \ud_tA_{ij}=-\left(A_{ik}A_{kj}-A_{kl}A_{lk}\frac{\delta_{ij}}{3}\right)+H_{ij}\;,
  \end{equation}
  where
  \[H_{ij}=-\left(\partial_i\partial_jp-\partial^2p\frac{\delta_{ij}}{3}\right)+\nu\partial^2A_{ij}\]
  is the (traceless) unclosed term for which a suitable
  model is required in order to express it in terms of $\bm{A}$ and other known quantities.
  Notice that, if one wants to investigate not the actual velocity field itself,
  but rather a coarse-grained version of it (e.g.\ in order to study velocity
  increments, as we will show at the end of section \ref{sec:lin},
  in the spirit of a large-eddy simulation (LES)), equation (\ref{vgt})
  remains valid, provided one includes the appropriate subgrid terms into $\bm{H}$ \cite{BO98,VTMK02}.

  The RE model takes $\bm{H}=\bm{0}$, i.e.\ it completely neglects
  the effect of viscosity and the anisotropic action of the (Eulerian)
  pressure Hessian, which are known to play an important role in the
  regularization and evolution of the turbulent dynamics. It is therefore
  not surprising that the RE model
  \[\ud_t\bm{A}=-\bm{A}^2+\frac{\tr{\bm{A}^2}}{3}\bm{I}\]
  leads to a finite time singularity.

  The RFD closure overcomes the blow-up by introducing four approximations
  and two time scales, namely a typical decorrelation (Kolmogorov) time $\tau$ and
  an integral (large-eddy turnover) time $T$. The closure is based on the mapping
  between the Eulerian position $\bm{x}$ (at time $t$) and an initial
  Lagrangian coordinate (or label) $\bm{X}$ (at some earlier time $t-\tau$). This map is
  invertible because of incompressibility, as the Jacobian has unit modulus.
  We briefly recall the four approximations used to obtain the closure.
  For more details, see \cite{CM06,CM07,CMBT08}.\\
  As a first step, when writing second derivatives by means of the
  chain differentiation rule, one neglects the spatial variations of the Jacobian:
  \begin{eqnarray*}
   \frac{\partial p}{\partial x_i}=\frac{\partial X_m}{\partial x_i}\frac{\partial p}{\partial X_m}&\Rightarrow&\frac{\partial^2p}{\partial x_i\partial x_j}\simeq\frac{\partial X_m}{\partial x_i}\frac{\partial X_n}{\partial x_j}\frac{\partial^2p}{\partial X_m\partial X_n}\;,\\
   \frac{\partial A_{ij}}{\partial x_k}=\frac{\partial X_m}{\partial x_k}\frac{\partial A_{ij}}{\partial X_m}&\Rightarrow&\frac{\nu\partial^2A_{ij}}{\partial x_k\partial x_k}\simeq\frac{\partial X_m}{\partial x_k}\frac{\partial X_n}{\partial x_k}\frac{\nu\partial^2A_{ij}}{\partial X_m\partial X_n}\;.
  \end{eqnarray*}
  Secondly, the \emph{Lagrangian} pressure Hessian is assumed isotropic
  (this is physically more meaningful than the corresponding
  approximation in RE about the isotropy of the \emph{Eulerian} Hessian):
  \[\frac{\partial^2p}{\partial X_m\partial X_n}\propto\delta_{mn}\]
  (the proportionality prefactor can be found through the Poisson equation).
  Moreover, one models the viscous term via an isotropic linear damping with the integral time scale $T$:
  \[\frac{\nu\partial^2A_{ij}}{\partial X_m\partial X_n}\simeq-\frac{A_{ij}}{T}\frac{\delta_{mn}}{3}\;.\]
  Finally, the crucial step is represented by the parameterization
  of the well-known Cauchy--Green tensor,
  \[C_{ij}\equiv\frac{\partial x_i}{\partial X_k}\frac{\partial x_j}{\partial X_k}\;,\]
  with an ``on-off approximation'':
  \begin{equation} \label{CGt}
   \bm{C}\simeq\ue^{\tau\bm{A}}\ue^{\tau\bm{A}^{\mathtt{T}}}\;,
  \end{equation}
  i.e., the actual slow decorrelation of $\bm{C}$ along the Lagrangian trajectory
  is replaced by a perfect correlation only during the Kolmogorov time scale $\tau$
  and by a complete decorrelation for longer times: in this way, the cumbersome,
  full time-ordered exponential reduces to a much simpler matrix exponential
  (here, the superscript $^{\mathtt{T}}$ denotes the transposed matrix).
  The Reynolds number (based on the integral scale) corresponds to
  $\mathrm{Re}=(\tau/T)^{-2}$, or in other words the Taylor-scale
  Reynolds number behaves as $\mathrm{Re}_{\lambda}\sim\tau^{-1}$.
  From (\ref{vgt}), using these approximations, and including a stochastic forcing term
  $\bm{F}$ to enable statistically stationary dynamics, the resulting system is:
  \begin{equation} \label{rfd}
   \ud_t\bm{A}=-\bm{A}^2+\bm{C}^{-1}~\frac{\tr\bm{A}^2}{\tr\bm{C}^{-1}}-\bm{A}~\frac{\tr\bm{C}^{-1}}{3T}+\bm{F}\;,
  \end{equation}
  with $\bm{C}$ given by (\ref{CGt}). Notice that the RE model
  corresponds to $\tau=0$, such that $\bm{C}=\bm{I}$, with $\bm{F}=\bm{0}$.
  In what follows, we will thus try to understand what happens when
  $\tau$ becomes smaller and smaller, a critical aspect of the RFD closure.

  Among related approaches to remove the blow-up
  behavior of RE, we can mention the tetrad model \cite{CPS99,NP05,NCP06,NPC07}, the Cauchy-Green models
  \cite{GP90a,JG03}, and a multi-scale coupling approach in the spirit of shell models \cite{BCMT07}.
  The RFD closure is attractive because of its relative simplicity: it is a system of only
  8 independent stochastic ordinary differential equations, based on a physically inspired connection to the pressure Hessian.
  However, unlike the shell model which allows to be extended to high Reynolds numbers by addition of shells,
  as already mentioned the RFD closure produces unrealistic predictions at increasing Reynolds numbers. In
  \cite{CM06,CM07}, $\tau\simeq0.06$ was the smallest value for which realistic results could be obtained. The analysis
  that follows explores the behavior at smaller $\tau$ in detail, in order to provide a basis for
  possible future improvements of the model.

 \section{Numerical investigation of the dynamics} \label{sec:num}

  In this section we present results based on numerical time integration
  of (\ref{rfd}), including the full matrix-exponential expression for $\bm{C}$.
  In terms of documenting results
  from the simulations, one can follow the evolution of the 9 (8 of which independent) components of
  $\bm{A}$ individually, e.g.\ in order to investigate their time correlations.
  One can also represent parts of the relevant dynamics in
  terms of the two invariants $R\equiv-\tr\bm{A}^3/3$ and $Q\equiv-\tr\bm{A}^2/2$.
  In the $R$-$Q$ plane the zero-discriminant curve $27R^2+4Q^3=0$ is
  usually called the Vieillefosse line (denoted in the plots as \textsf{Vline}): the RE model is known
  to diverge on its right-bottom part.

  As numerical procedure we adopt a standard fourth-order
  Runge--Kutta scheme for the time integration, with time step $\Delta t$ small enough
  (always $\Delta t\le10^{-3}\tau$, i.e.\ at least one thousandth of the shortest relevant time scale).
  As in \cite{CM06,CM07}, we perform a non-dimensionalization of the equations with $T$, thus the only remaining
  significant parameter is $\tau$ ($\sim\mathrm{Re}^{-1/2}$).
  We run the code for a total time $\ge10^3$ (i.e.\ at least one thousand large-eddy turnover times), in
  order to accumulate well-converged statistics. As a first test, we integrated the equations
  starting from several initial conditions without forcing ($\bm{F}=\bm{0}$),
  in order to check that the dynamics properly decay, converging to the origin.

  In order to obtain stationary statistics, the tensorial forcing term $\bm{F}$ is introduced
  on the RHS of (\ref{rfd}). Following arguments given in
  \cite{CM06,CM07}, this term mimics the effects of neighboring small-scale eddies
  that affect the velocity gradient in its Lagrangian evolution.
  We performed some preliminary tests by using a white-in-time
  random signal (following \cite{CM06}), but then we chose to focus on
  a Gaussian, smooth noise, with a finite time correlation $\theta$. This time scale will be chosen of the order
  of $\tau$ itself, because this is the time scale at which the
  interactions with the other eddies are believed to take place.
  A comparison between the two kinds of noise, and thus on the
  influence of the time correlation, will be made in subsection \ref{sec:timecorr},
  where various values of $\theta$ (namely
  $\theta=\tau,\ \tau/2,\ \tau/5,\ \tau/10$) will be considered.
  In appendix \ref{sec:app} we discuss some details of the forcing.

  From (\ref{CGt}) it appears that the tensor $\bm{C}$ will become important once
  $\bm{A}$ should have order of magnitude $\sim\tau^{-1}$. The quadratic terms on the
  RHS of (\ref{rfd}) then suggest that $\bm{A}$ should vary on a characteristic time scale of order $\tau$.
  Thus, as a first guess for the scaling of the amplitude of the noise, it was chosen of the order of $\tau^{-2}$
  so that it is comparable to the magnitude of $\bm{A}^2$. However, for
  $\tau\sim10^{-2}$ and smaller, this turned out to be a too strong a forcing amplitude: the forcing overwhelmed the quadratic
  self-interaction term, resulting in nearly Gaussian statistics for $\bm{A}$.
  Therefore, a more careful procedure must be followed to
  prescribe the amplitude of the forcing. We begin by taking the value $\tau=10^{-1}$ as a
  baseline case (this value was the baseline value also for the prior studies \cite{CM06,CM07,CMBT08}). For
  this reference case, in \cite{CM06,CMBT08} an amplitude of the forcing $\bm{F}=O(1)$ was used,
  along with a compensation with the integration time step,
  due to the white-noise character of the forcing used there.
  Let us denote the actual amplitude of the forcing $\mathcal{F}$.
  Equation (\ref{oup}) in the appendix shows how the forcing is
  generated as function of $\mathcal{F}$.
  Since we wish to compare our results with the ones of \cite{CM06,CMBT08},
  we also adopt $\mathcal{F}=1$ for this reference case.
  As shown in appendix \ref{sec:app}, $\mathcal{F}$ is used in
  (\ref{oup}) as amplitude of the diffusion term.
  Note that $F_{12}^{\mathrm{rms}}=\mathcal{F}$.

\begin{figure}
   \centering
   \includegraphics[scale=.6]{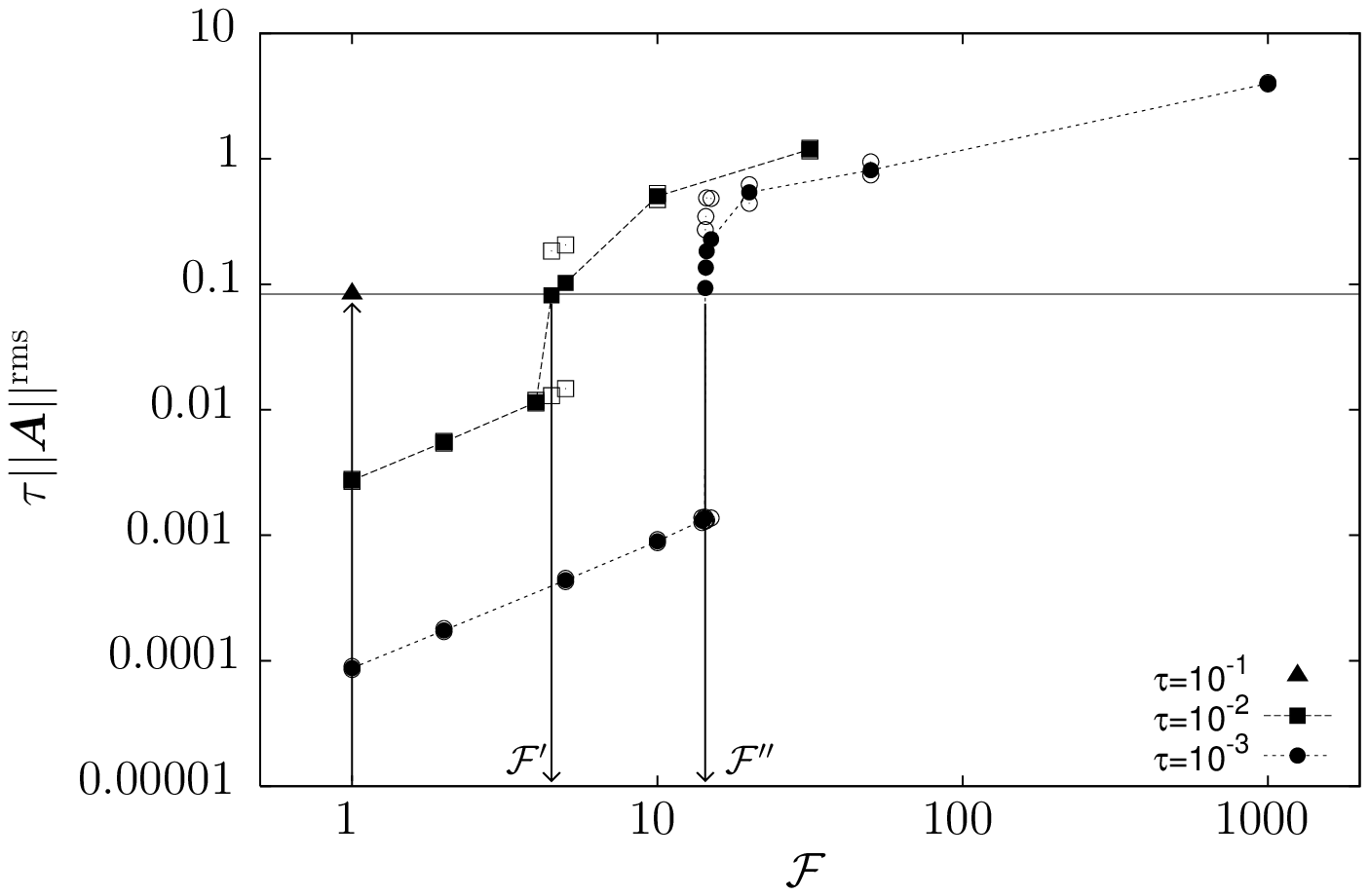}
   \caption{Response behavior of $||\bm{A}||^{\mathrm{rms}}$, multiplied by $\tau$, as a function of the typical noise amplitude $\mathcal{F}$,
    for three different values of $\tau$. The two (approximate) intercepts with the horizontal line passing through the point representative
    of the reference case $\tau=10^{-1}$ (for which one only value is clearly plotted, corresponding to $\mathcal{F}=1$)
    represent the correct values of $\mathcal{F}$ which will be used in what follows for the cases $\tau=10^{-2}$ and $=10^{-3}$,
    i.e.\ $\mathcal{F}'$ and $\mathcal{F}''$ respectively. The arrows have been introduced to help the understanding of the procedure.
    For every point (full squares and circles), the corresponding couple of empty symbols (above and below it,
    in some cases overshadowed) indicates the minimum and maximum values
    obtained by subdividing the whole numerical run into 10 equal sub-intervals, and is thus an indication of the
    statistical uncertainty appearing in the ranges where the increase in amplitudes is very steep.}
   \label{fig:points}
  \end{figure}

  As a next step, we consider increasing $\mathrm{Re}$, and we know that the strength of
  the forcing (i.e.\ of the action of neighboring eddies) should
  increase, but at a rate smaller than $\sim\tau^{-2}$, as already
  explained. An important criterion that the dynamics should follow when increasing $\mathrm{Re}$ is that, for energy
  dissipation to remain constant at increasing $\mathrm{Re}$, one should observe that
  $||\bm{A}||^{\mathrm{rms}}\equiv\sqrt{\langle A_{ij}^2\rangle}\sim\tau^{-1}$, when $\tau$ is interpreted as the Kolmogorov time scale \cite{CM06}.
  Hence, the forcing may be adjusted to generate approximately such scaling.
  In figure \ref{fig:points} we plot, in logarithmic scale, the
  response behavior of $||\bm{A}||^{\mathrm{rms}}$, multiplied by
  $\tau$, as a function of $\mathcal{F}$, as obtained from a large number of simulations.
  We observe that $\tau||\bm{A}||^{\mathrm{rms}}$ increases monotonically
  until a certain forcing strength, and then a sudden increase
  in amplitude is observed. Beyond this transition,
  the increase continues less dramatically.
  Following the energy-dissipation-based criterion
  (of aiming to reproduce $\tau||\bm{A}||^{\mathrm{rms}}\simeq0.08$),
  the values $\mathcal{F}'=4.5$ and $\mathcal{F}''=14.35$, obtained by
  finding the approximate intercepts with the horizontal line
  drawn through the reference-case point, represent the correct
  amplitudes to be used for $\tau=10^{-2}$ and $=10^{-3}$, respectively.

  \subsection{Baseline results}

\begin{figure}
   \centering
   \includegraphics[scale=.6]{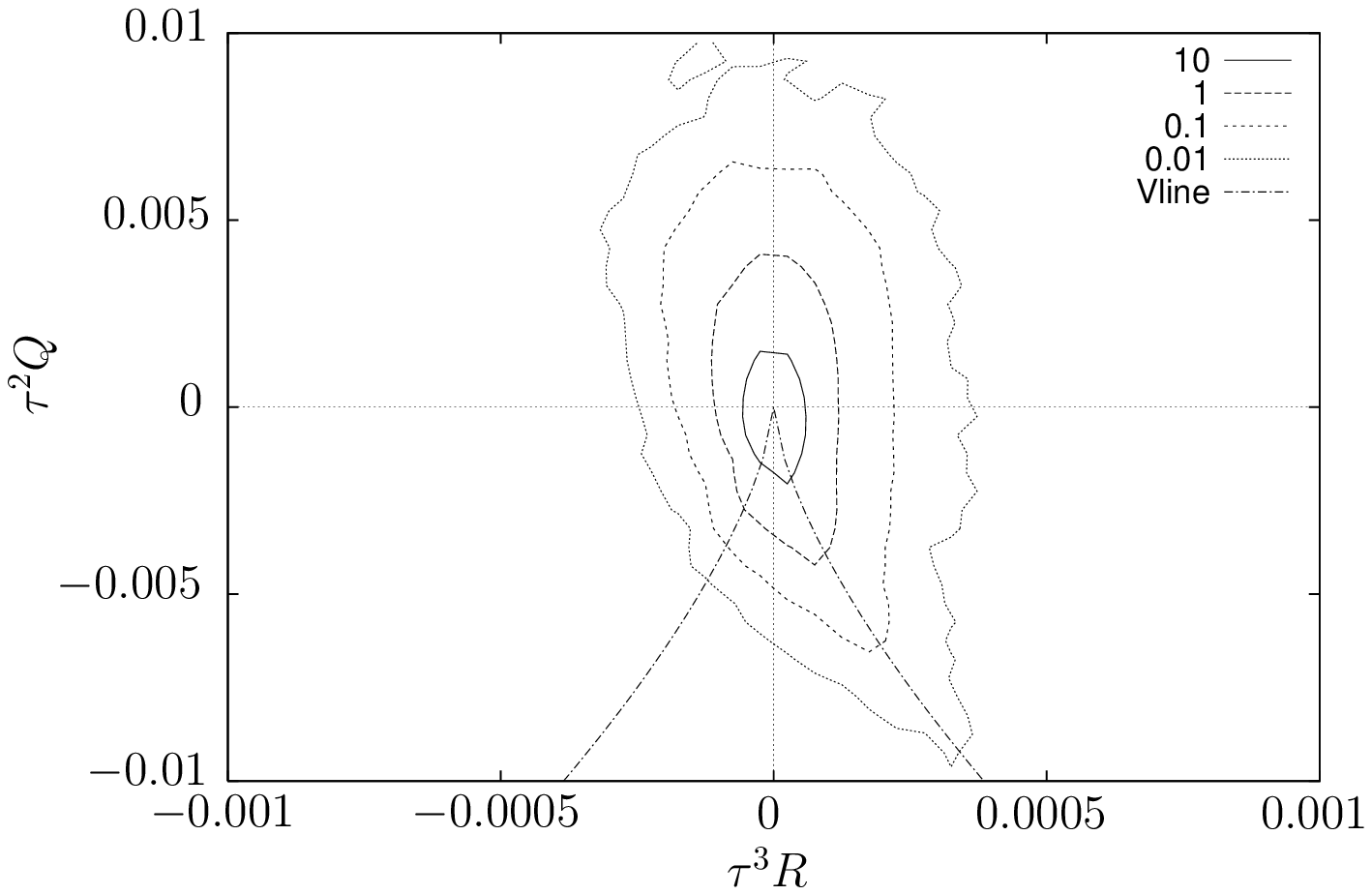}
   \caption{Joint PDF of $R$-$Q$ obtained from the RFD closure, using $\tau=10^{-1}$
    and forcing with time correlation $\theta=\tau$ and typical amplitude $\mathcal{F}=1$.
    Note that $P(R,Q)$ is plotted as a function of scaled values $\tau^3R$ and $\tau^2Q$, instead of $R$ and $Q$.}
   \label{fig:m-1}
  \end{figure}

\begin{figure}
   \centering
   \includegraphics[scale=.6]{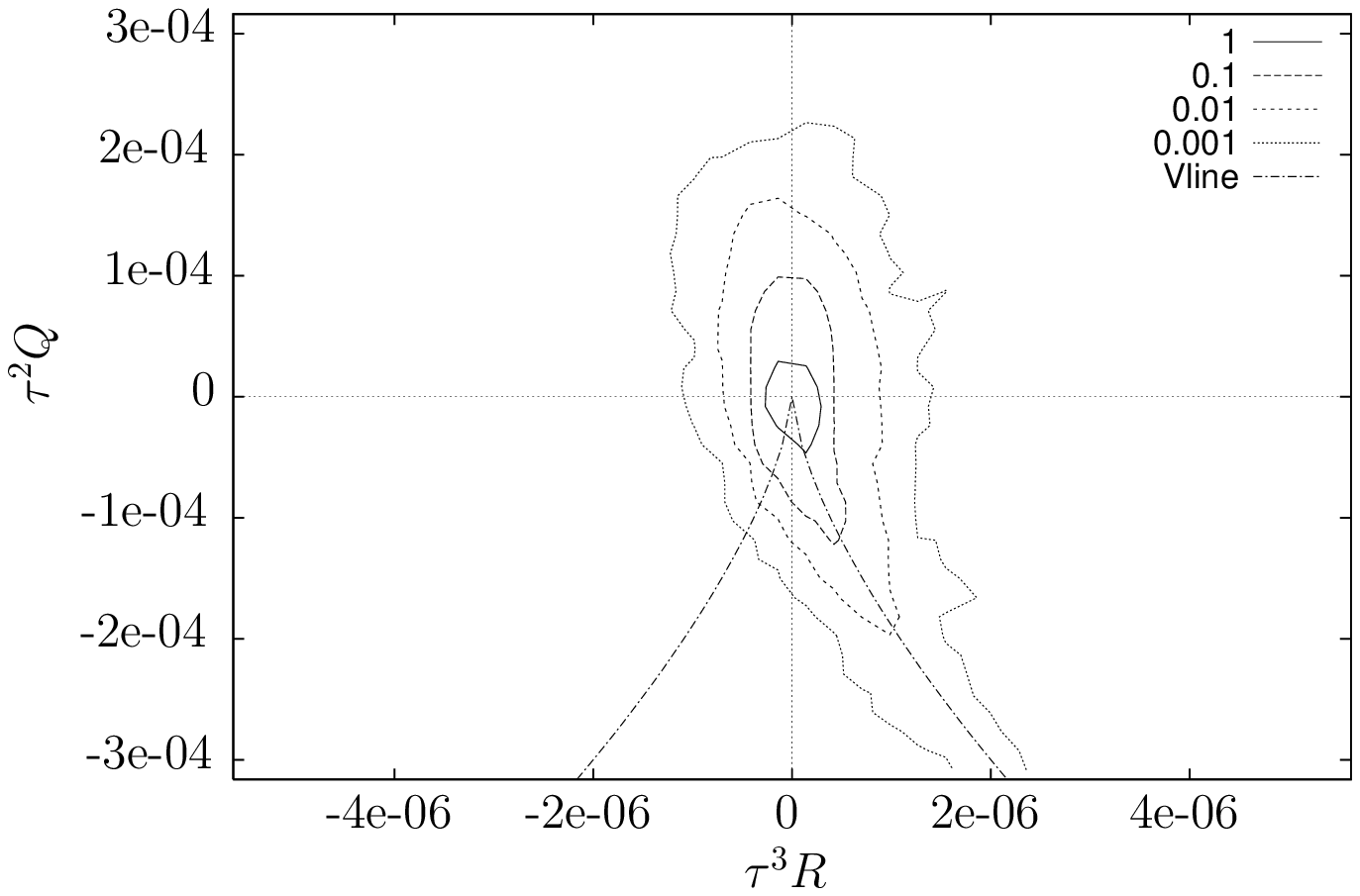}
   \caption{Same as in figure \ref{fig:m-1}, but for $\tau=10^{-2}$ and $\mathcal{F}=4.5$.}
   \label{fig:m-2}
  \end{figure}

\begin{figure}
   \centering
   \includegraphics[scale=.6]{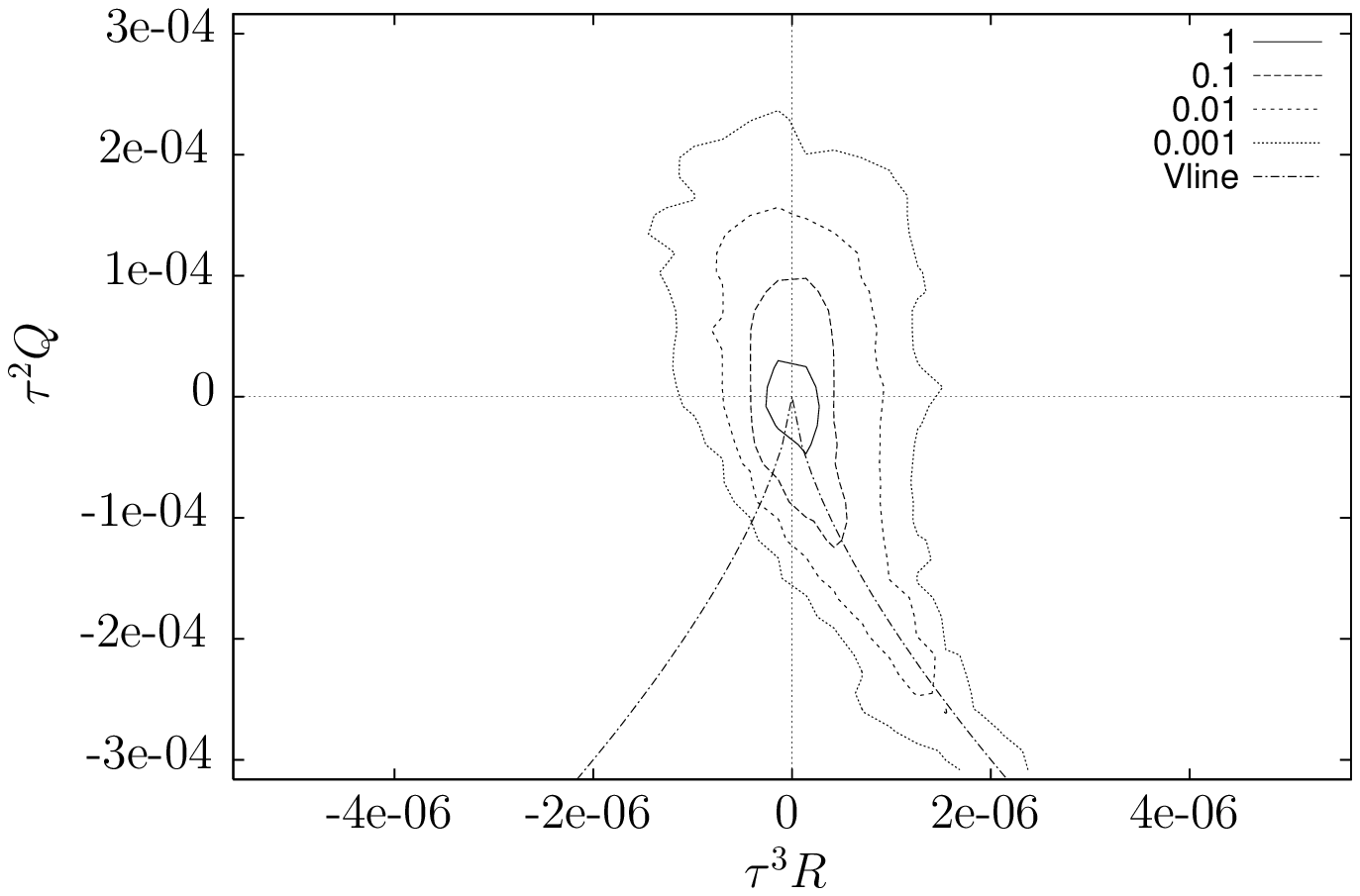}
   \caption{Same as in figure \ref{fig:m-1}, but for $\tau=10^{-3}$ and $\mathcal{F}=14.35$.}
   \label{fig:m-3}
  \end{figure}

  The baseline case of $\tau=10^{-1}$ is computed, using a Gaussian random forcing with time correlation $\theta=\tau$
  and with forcing parameter $\mathcal{F}=1$. The model yields the probability density function
  for $Q$ and $R$ shown in figure \ref{fig:m-1}, which agrees well with \cite{CM06}.
  It is worth mentioning that, differently from \cite{CM06}, here we multiply $Q$ and $R$ by (the
  appropriate power of) $\tau$, rather than normalizing them with the trace of $[(\bm{A}+\bm{A}^{\mathtt{T}})/2]^2$.\\
  Next, we consider the case of $\tau=10^{-2}$ (with $\mathcal{F}=\mathcal{F}'$,
  such that $||\bm{A}||^{\mathrm{rms}}$ becomes 10 times bigger than in the previous case)
  and show the results in figure \ref{fig:m-2}. The shape of the
  PDF is reasonable, but one should notice that the values of $R$
  and $Q$ do not scale with factors $10^3$ and $10^2$ with respect
  to figure \ref{fig:m-1}, as one would expect from a simple
  dimensional analysis. This is a clear indication of the presence
  of a very intermittent behavior, as we will show in subsection \ref{sec:timecorr}.\\
  This observation is confirmed further when plotting the case $\tau=10^{-3}$
  (figure \ref{fig:m-3}), for which the picture is very similar to
  the latter case, in the sense that the rescaling of $R$ and $Q$
  is much smaller than expected.

\begin{figure}
   \centering
   \includegraphics[scale=.6]{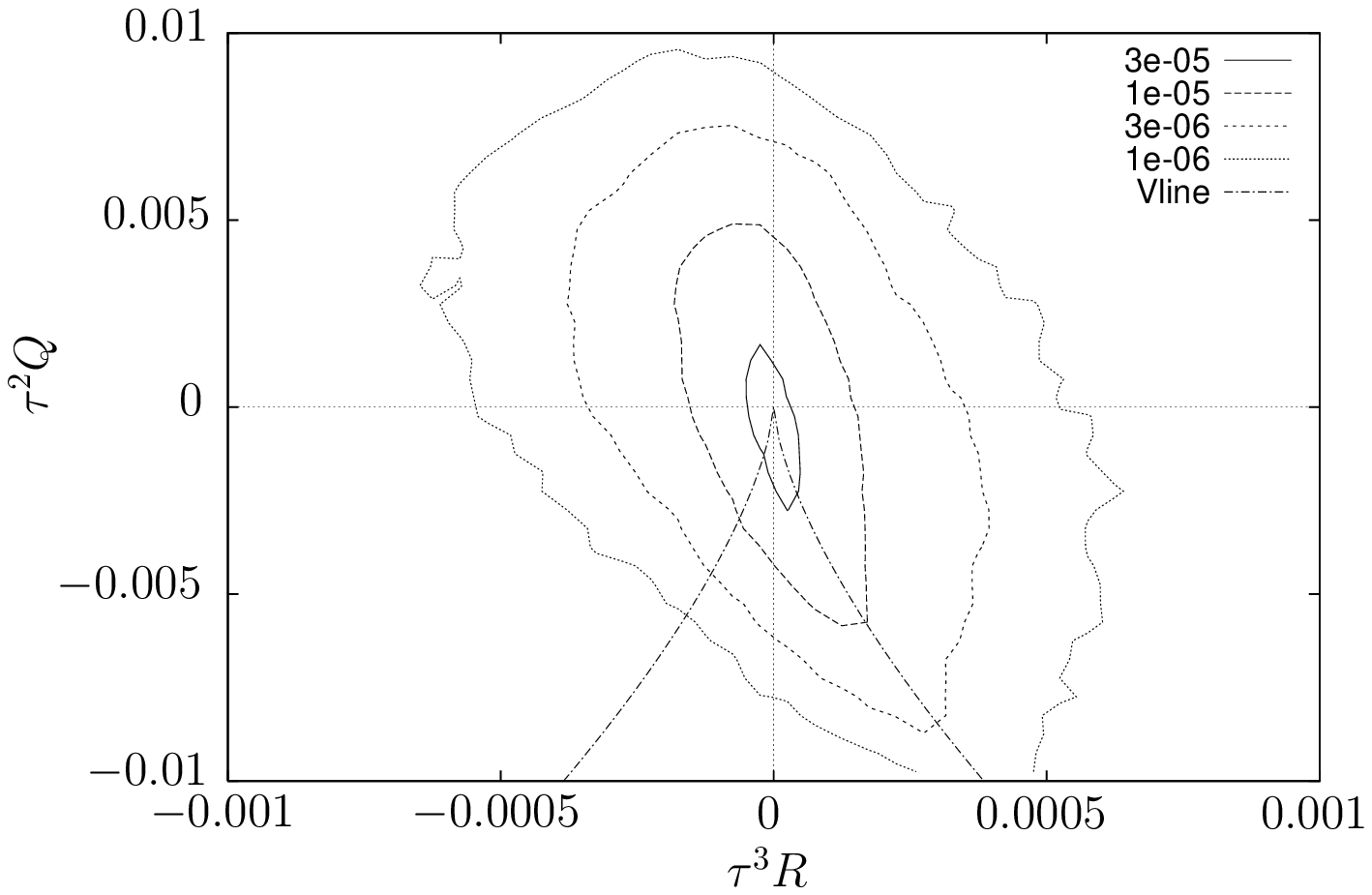}
   \caption{Same as in figure \ref{fig:m-2}, but for $\mathcal{F}=31.6$.}
   \label{fig:m2}
  \end{figure}

\begin{figure}
   \centering
   \includegraphics[scale=.6]{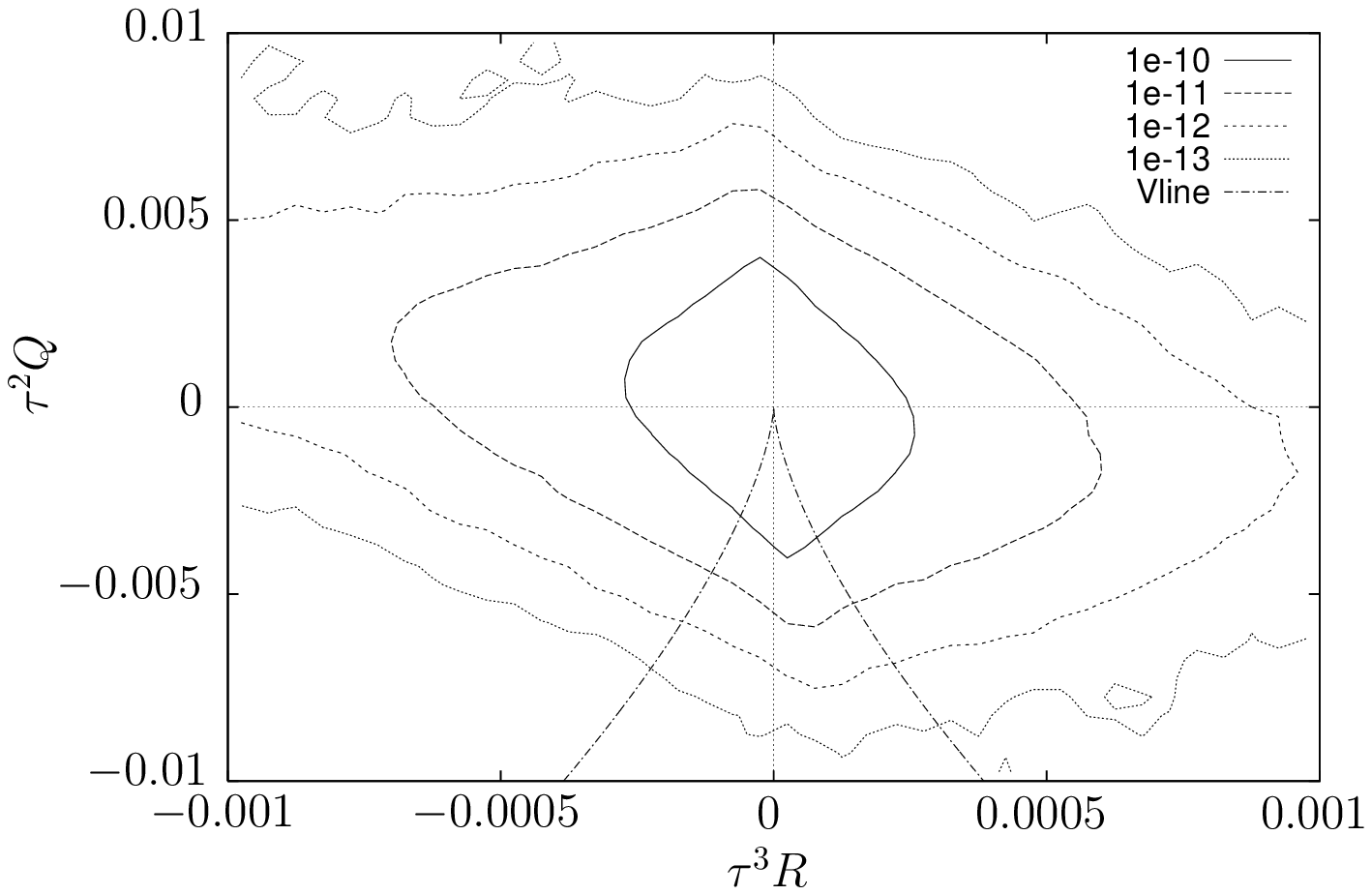}
   \caption{Same as in figure \ref{fig:m-3}, but for $\mathcal{F}=1000$.}
   \label{fig:m3}
  \end{figure}

  One could therefore wonder what happens when the amplitude of
  the forcing is chosen in a way such as to give results whose
  orders of magnitude of $Q$ and $R$ are
  consistent with the expected $Q\sim\tau^{-2}$ and $R\sim\tau^{-3}$ scaling. In
  other words, for $\tau=10^{-2}$ we momentarily impose a different value
  $\mathcal{F}=31.6$ ($\gg\mathcal{F}'$) such that, passing from figure \ref{fig:m-1}
  to figure \ref{fig:m2}, a
  rescaling of $\tau$ by one order of magnitude
  ($10^{-1}$) brings about an approximate rescaling of the quadratic and
  cubic quantities, $Q$ and $R$, by factors $10^2$ and $10^3$ respectively.
  The results are still reasonable in terms of the shape of the PDF.\\
  However, at $\tau=10^{-3}$ (figure \ref{fig:m3}), when using the same criterion
  (which implies $\mathcal{F}=1000$, $\gg\mathcal{F}''$) the model yields the results shown in figure \ref{fig:m3}.
  The resulting PDF displays an unrealistic diamond-like
  shape skewed towards the second and fourth quadrants. Further modifying the amplitude of the noise does
  not bring about better results for this last case.

  Besides testing other forcing amplitudes, we have also tried other possible forms for the forcing,
  including various multiplicative noises. Among them, we can mention:
  $\ud_t\bm{A}=\ldots+(\bm{F}\cdot\bm{A})^{\mathtt{D}}$,
  $\ud_t\bm{A}=\ldots+\bm{F}||\bm{A}||^2$,
  $\ud_t\bm{A}=\ldots+\bm{F}(1+||\bm{A}||^2)$
  and $\ud_t\bm{A}=\ldots+\bm{F}(1+Q)$,
  where the superscript $^{\mathtt{D}}$ identifies the deviatoric part
  of the tensor (i.e.\ the tensor minus its trace times $\bm{I}/3$).
  However, none of them lead to physically meaningful results for small $\tau$.

  Analogously, following \cite{CM07}, we tried to impose not a
  fixed, overall dissipative time scale, but rather a variable,
  local one, $\tau(t)\propto[\tr(\bm{A}+\bm{A}^{\mathtt{T}})^2]^{-1/2}$.
  However, also this approach did not allow us to obtain better,
  or more physical, results for cases corresponding to higher Reynolds numbers.

  \subsection{Time correlation structure of velocity gradients}  \label{sec:timecorr}

  In this section, we document time series, autocorrelation functions and time scales of
  velocity gradient tensor elements as predicted by the model. Several options of the forcing term are considered.
  Specifically, we vary the time correlation $\theta$ of the forcing. Five cases are considered:
  the baseline case with $\theta=\tau$ itself,
  the white-noise case (i.e.\ $\theta=0$),
  and three more finite-correlated instances with intermediate
  time correlations of $\theta=\tau/2$, $\tau/5$ and $\tau/10$.
  Each case needs a modified amplitude $\mathcal{F}$ in order to try to maintain the
  amplitude of the velocity gradient tensor comparable, as $\theta$ is varied.
  Inspired by the behavior of a forced linear system, we use the rescaling in which $\mathcal{F}$
  is replaced by $\mathcal{F}|_{\theta=\tau}\sqrt{\tau/\theta}$ for the finite-correlated forcings
  and by $\mathcal{F}|_{\theta=\tau}\sqrt{\tau/\Delta t}$ for the white-noise case.
  It is observed that this rescaling indeed maintains the typical
  amplitudes of $\bm{A}$ unchanged even as $\theta$ is varied.

\begin{figure}
   \centering
   \includegraphics[scale=.6]{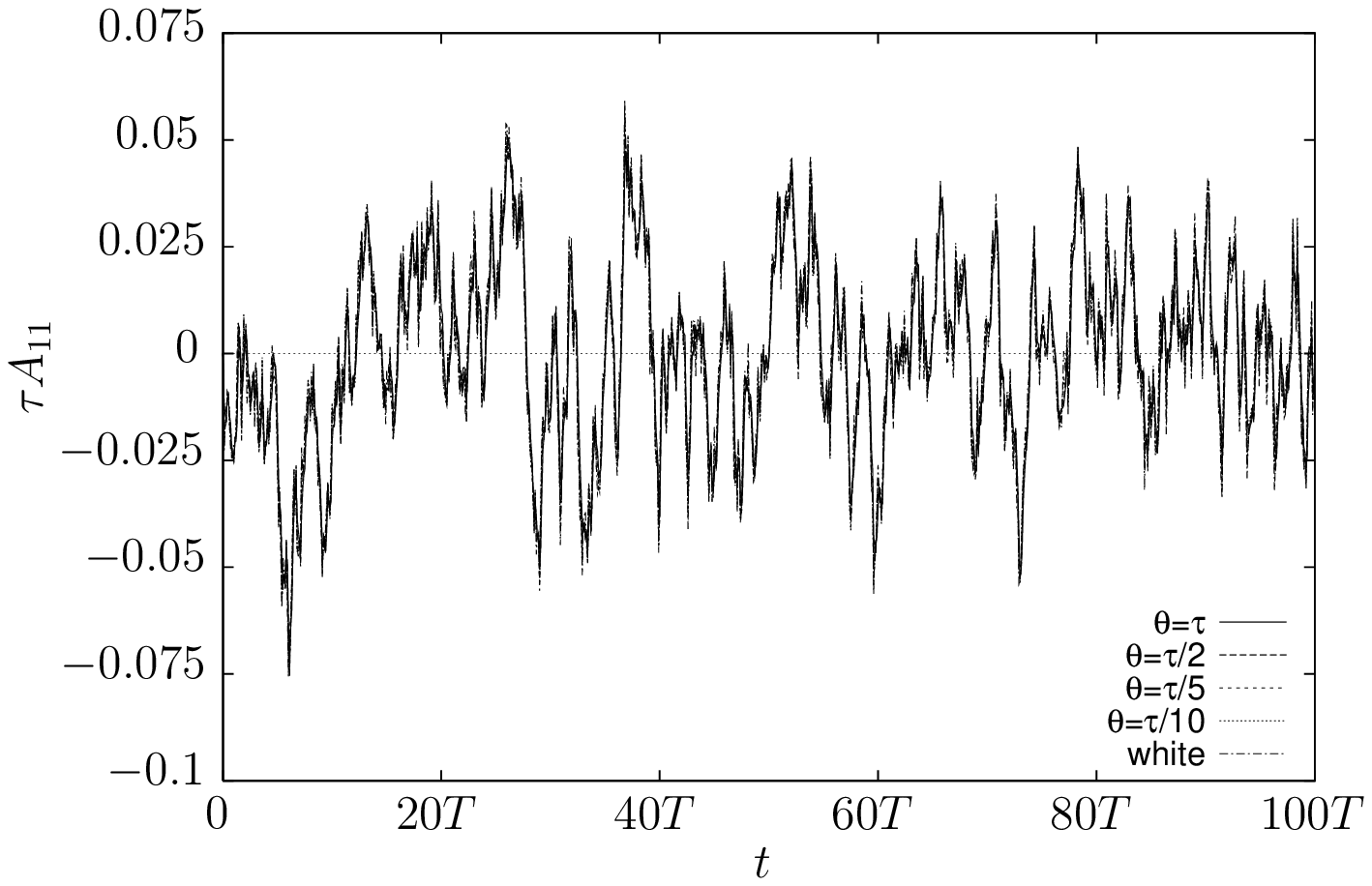}
   \includegraphics[scale=.6]{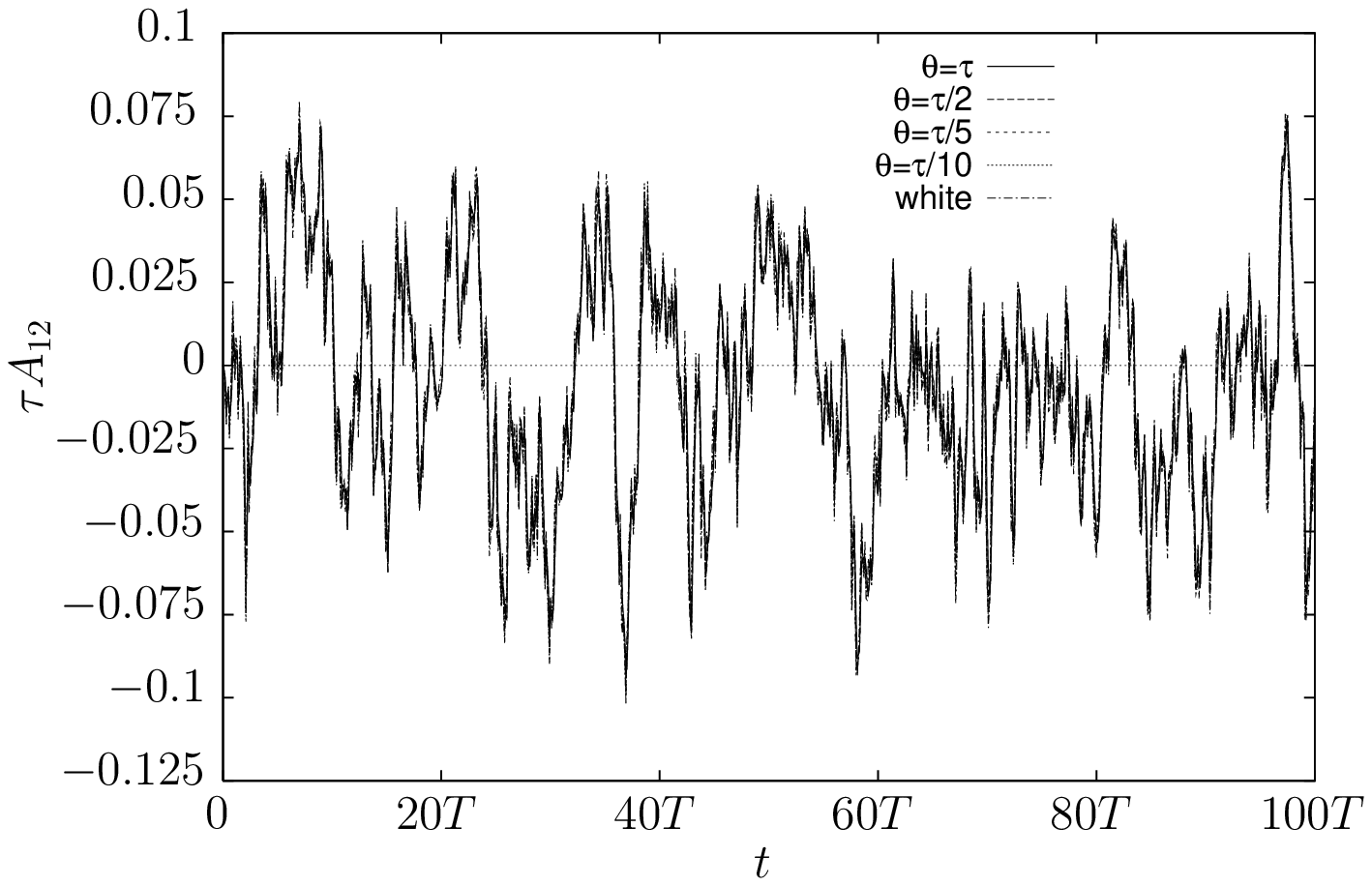}
   \caption{Time signal of one longitudinal (top panel) and one transverse (bottom panel) component of the simulated velocity gradient tensor, multiplied by $\tau$, at $\tau=10^{-1}$ and $\mathcal{F}=1$. Five lines are shown, corresponding to different time correlations of the forcing. Note that the plotted time window has been chosen as $100$ (i.e.\ $100T$ in units), so that $\tau$ corresponds to one thousandth of the total horizontal extension.}
   \label{fig:s-1}
  \end{figure}

\begin{figure}
   \centering
   \includegraphics[scale=.6]{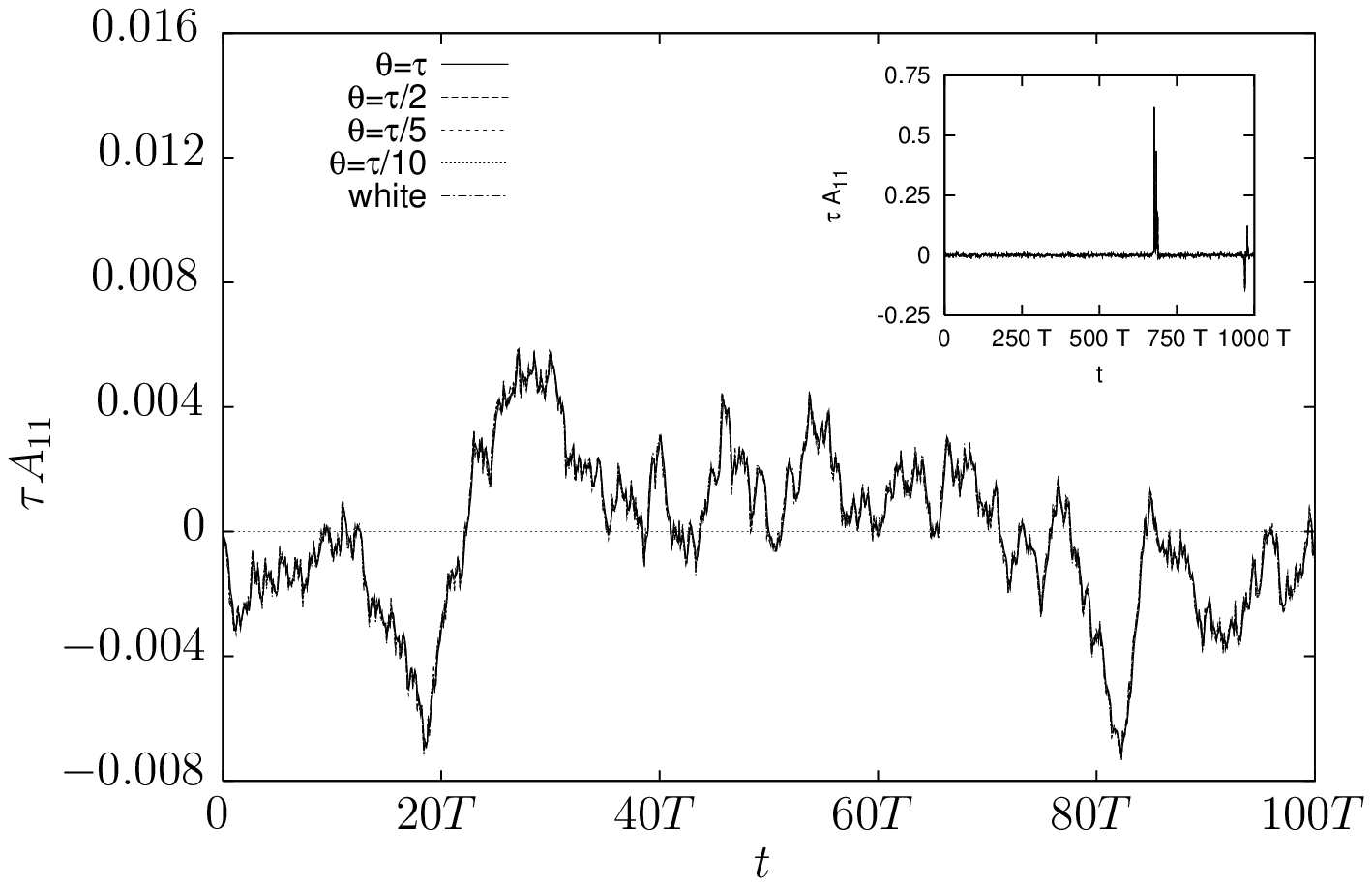}
   \caption{Same as in the top panel of figure \ref{fig:s-1} but with $\tau=10^{-2}$ and $\mathcal{F}=\mathcal{F}'$. Note that $\tau$ corresponds to one tenth of thousandth of the total horizontal extension. The insert shows the results of the full, longer run, whose first part is reproduced in the outer plot.}
   \label{fig:s-2}
  \end{figure}

\begin{figure}
   \centering
   \includegraphics[scale=.6]{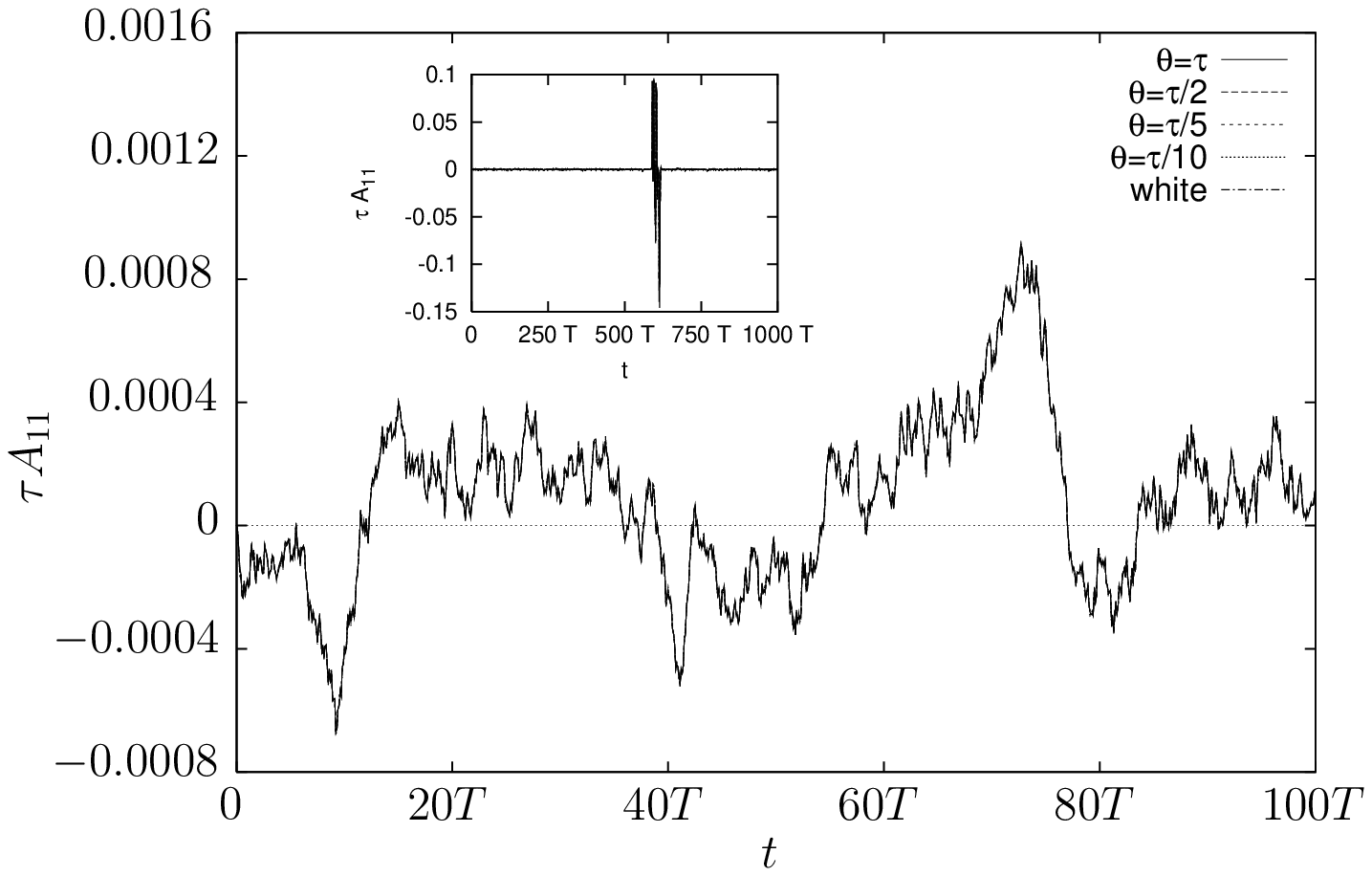}
   \caption{Same as in figure \ref{fig:s-2} but with $\tau=10^{-3}$ and $\mathcal{F}=\mathcal{F}''$. Note that $\tau$ corresponds to one hundredth of thousandth of the total horizontal extension.}
   \label{fig:s-3}
  \end{figure}

  In figure \ref{fig:s-1}, we plot the signal, as a function of time, of one longitudinal ($A_{11}$) and one
  transverse ($A_{12}$) component of the velocity gradient tensor,
  at $\tau=10^{-1}$. The same analysis is repeated in figures
  \ref{fig:s-2}--\ref{fig:s-3}, at $=10^{-2}$ and $10^{-3}$ respectively,
  only for the longitudinal component.
  In all cases, there seem to be essentially no differences between the
  five evolutions plotted, which correspond to different
  time correlations $\theta$ of the forcing.
  Note from the inserts (corresponding to longer, full numerical runs)
  that the cases for $\tau=10^{-2}$ and $10^{-3}$ display some very
  infrequent but violent fluctuations, which are absent for the baseline case
  $\tau=10^{-1}$. These fluctuations are responsible for
  the sudden increase in amplitudes seen in figure
  \ref{fig:points}. A detailed study of the dynamics during such
  intermittent excursions is left for a future study.

\begin{figure}
   \centering
   \includegraphics[scale=.63]{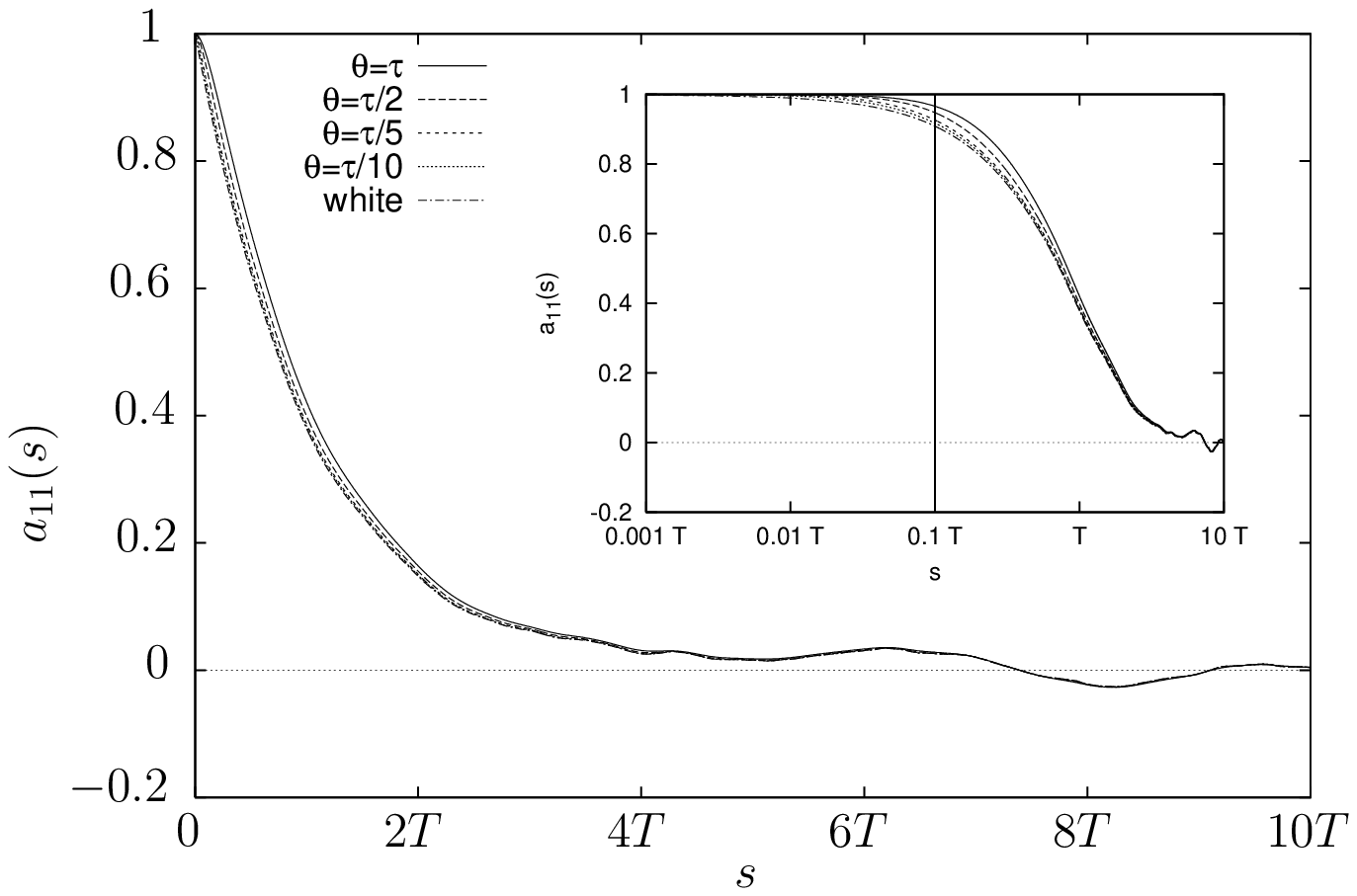}
   \includegraphics[scale=.63]{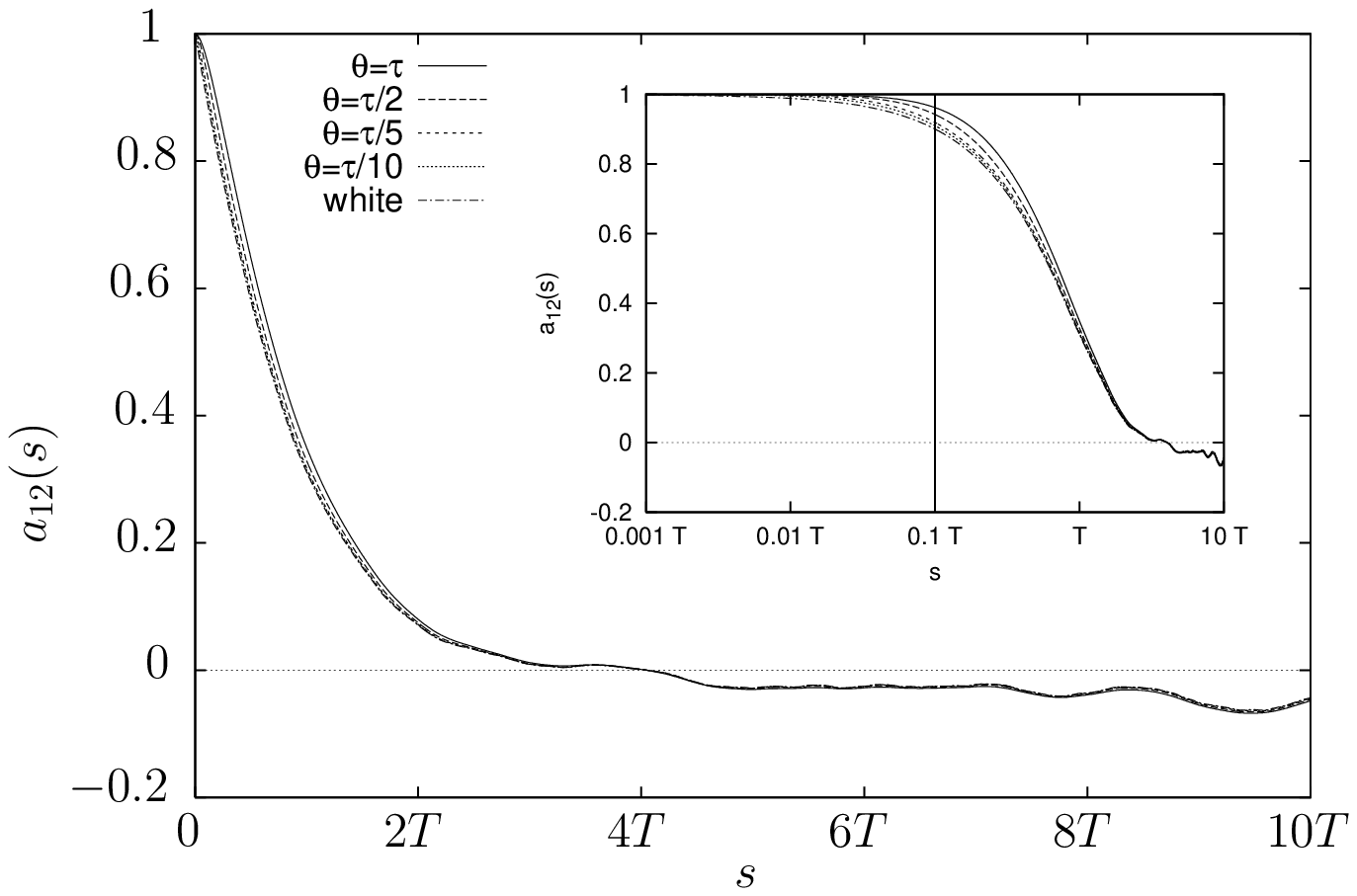}
   \caption{Time autocorrelation of one longitudinal, $a_{11}(s)$ (top panel), and one transverse, $a_{12}(s)$ (bottom panel), component of the velocity gradient tensor, at $\tau=10^{-1}$ and for five different values of $\theta$. Notice that, in the linear scale (outer plot), $\tau$ corresponds to one hundredth of the total horizontal extension $10T$. In the insert, the axis of the abscissae is represented in logarithmic scale, and the vertical line indicates $s=\tau$.}
   \label{fig:c-1}
  \end{figure}

\begin{figure}
   \centering
   \includegraphics[scale=.63]{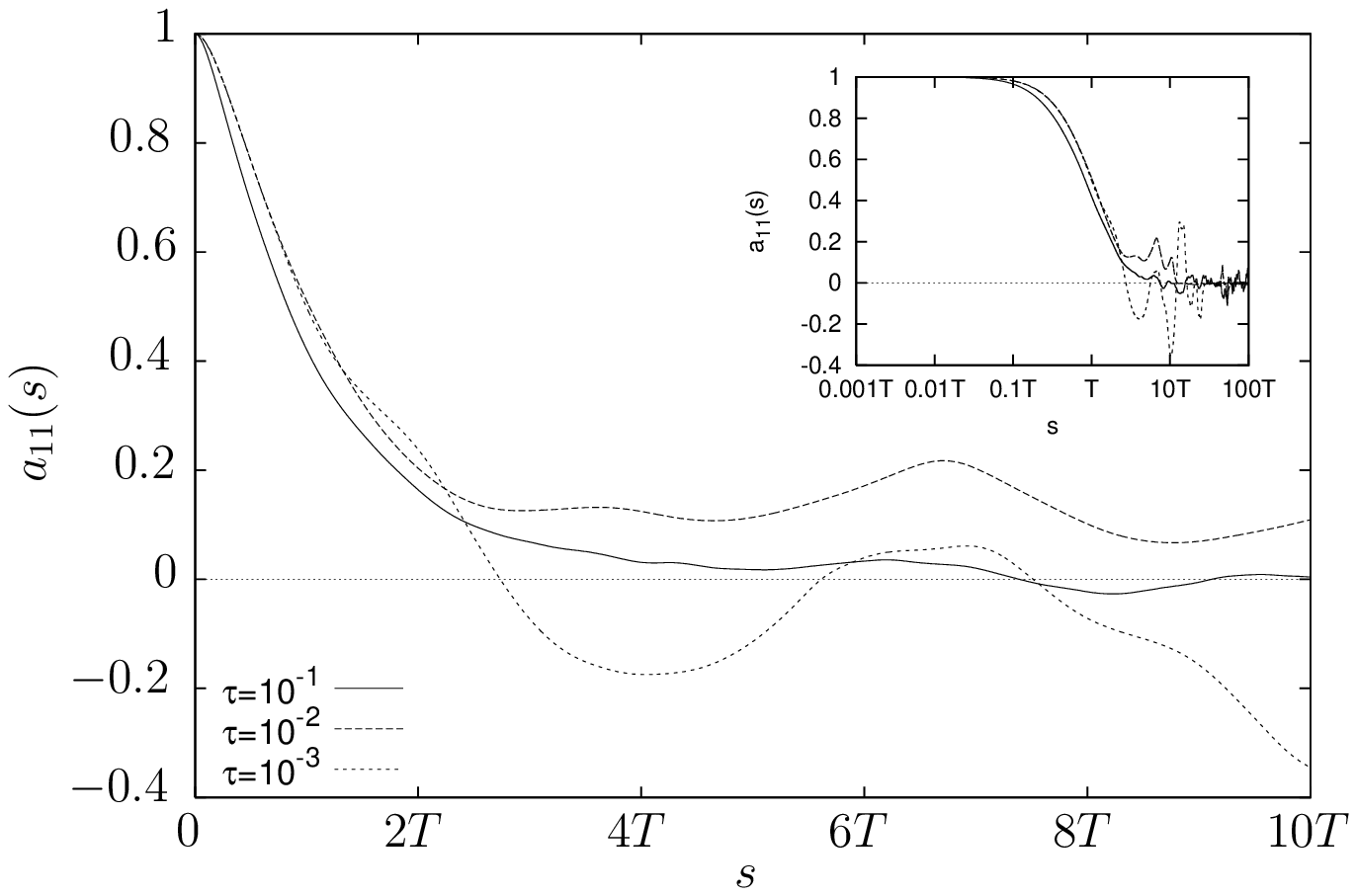}
   \caption{Time autocorrelation of one longitudinal component of the velocity gradient tensor, $a_{11}(s)$, for three different values of $\tau$. In the insert, the axis of the abscissae is represented in logarithmic scale.}
   \label{fig:comp}
  \end{figure}

  Next, the time correlations of $\bm{A}$ are quantified,
  for simulations using the same five forcing cases.
  In figure \ref{fig:c-1}, for both one longitudinal
  and one transverse component, we plot the autocorrelation function
  \[a_{ij}(s)\equiv\frac{\langle A_{ij}(t)A_{ij}(t+s)\rangle-\langle A_{ij}(t)\rangle^2}{\langle A_{ij}^2(t)\rangle-\langle A_{ij}(t)\rangle^2}\]
  (no summation implied), where the temporal average
  has been performed on our entire numerical run.
  Figure \ref{fig:comp} shows the results
  for $\tau=10^{-2}$ and $10^{-3}$ as well for long iterations (only the three cases $\theta=\tau$ are plotted).
  A striking observation is that the measured correlation time scale from the autocorrelation function is quite large, and in both
  cases reaches $0.2$ only at a time scale $s\sim T$ ($=1$), significantly larger than $\tau$. More surprisingly, when decreasing the
  imposed time scale $\tau$ (which corresponds to an assumed decorrelation time in the RFD closure), the results
  from the simulations certainly do not show a concomitant decrease in the predicted autocorrelation function.
  Also for $\tau=10^{-2}$ and $10^{-3}$, we observe that the autocorrelation only approaches zero for $s\sim T$ and not $\tau$.

 \section{Power-series expansion of the matrix exponential} \label{sec:exp}

  One major effect of the RFD closure for the pressure Hessian is to oppose the generation of finite-time singularities
  that occur in the RE equation. Hence, at least in parts of the $R$-$Q$ plane, the closure acts as a regularization or damping effect.
  Detailed comparisons with direct numerical simulations (DNS) in \cite{CMBT08} confirm this effect
  using conditional averaging analysis. In order to study this issue by inspecting the closure terms directly,
  it is convenient to introduce a simplification of the matrix exponential using expansions.
  The definition of the matrix exponential is used to obtain a power-series expansion:
  \[\ue^{\tau\bm{A}}\equiv\sum_{n=0}^{\infty}\frac{1}{n!}(\tau\bm{A})^n=\bm{I}+\tau\bm{A}+\frac{1}{2}\tau^2\bm{A}^2+\frac{1}{6}\tau^3\bm{A}^3+\ldots\;.\]
  Replacing in (\ref{CGt}) and (\ref{rfd}), one obtains:
  \begin{eqnarray} \label{expan}
   \ud_t\bm{A}&=&-(\bm{A}^2)^{\mathtt{D}}+\frac{\tr\bm{A}^2}{3}\left[\frac{\bm{A}^2+(\bm{A}^{\mathtt{T}})^2}{2}+\bm{A}^{\mathtt{T}}\bm{A}\right]^{\mathtt{D}}\tau^2\nonumber\\
   &&\!\!-\frac{\tr\bm{A}^2}{3}\left[\frac{\bm{A}^3+(\bm{A}^{\mathtt{T}})^3}{6}+\frac{\bm{A}^{\mathtt{T}}\bm{A}^2+(\bm{A}^{\mathtt{T}})^2\bm{A}}{2}\right.\nonumber\\
   &&\left.\ -(\bm{A}+\bm{A}^{\mathtt{T}})\frac{\tr\bm{A}^2+\tr\bm{A}\bm{A}^{\mathtt{T}}}{3}\right]^{\mathtt{D}}\tau^3\nonumber\\
   &&\!\!-\left[\frac{\bm{A}}{T}+(\bm{A}+\bm{A}^{\mathtt{T}})\frac{\tr\bm{A}^2}{3}\tau+\frac{\bm{A}}{T}\frac{\tr(\bm{A}+\bm{A}^{\mathtt{T}})^2}{6}\tau^2\right.\nonumber\\
   &&\left.\ -\frac{\bm{A}}{T}\frac{\tr(\bm{A}+\bm{A}^{\mathtt{T}})^3}{18}\tau^3\right]+O(\tau^4)\;.
  \end{eqnarray}

  It is also interesting to recast (\ref{expan}) by separating it
  into its symmetric and antisymmetric components,
  $\bm{S}\equiv(\bm{A}+\bm{A}^{\mathtt{T}})/2$
  and $\bm{\Omega}\equiv\bm{A}-\bm{S}$. We get:
  \begin{eqnarray}
   \ud_t\bm{S}&=&-\left(\bm{S}^2+\bm{\Omega}^2\right)^{\mathtt{D}}-\frac{2}{3}Q\left[2(\bm{S}^2)^{\mathtt{D}}+\bm{S}\bm{\Omega}-\bm{\Omega}\bm{S}\right]\tau^2\nonumber\\
   &&\!\!+\frac{2}{9}Q\left[4(\bm{S}^3)^{\mathtt{D}}-2\bm{\Omega}\bm{S}\bm{\Omega}+\bm{S}\bm{\Omega}^2+\bm{\Omega}^2\bm{S}\right.\nonumber\\
   &&\left.\ +3\bm{S}^2\bm{\Omega}-3\bm{\Omega}\bm{S}^2+8Q_{\bm{S}}\bm{S}\right]^{\mathtt{D}}\tau^3\nonumber\\
   &&\!\!-\frac{\bm{S}}{T}\left\{1-\frac{4}{3}TQ\tau-\frac{4}{3}Q_{\bm{S}}\tau^2+\frac{4}{3}R_{\bm{S}}\tau^3\right\}\nonumber\\
   &&+O(\tau^4)\;,
   \label{dotS}\\&&\nonumber\\\label{dotO}
   \ud_t\bm{\Omega}&=&-(\bm{S}\bm{\Omega}+\bm{\Omega}\bm{S})-\frac{\bm{\Omega}}{T}\left\{1-\frac{4}{3}Q_{\bm{S}}\tau^2+\frac{4}{3}R_{\bm{S}}\tau^3\right\}\nonumber\\
   &&+O(\tau^4)\;.
  \end{eqnarray}
  Here, $Q_{\bm{S}}\equiv-\tr\bm{S}^2/2$ and $R_{\bm{S}}\equiv-\tr\bm{S}^3/3$
  are the second and third invariants of $\bm{S}$
  (its first invariant, i.e.\ its trace, is identically zero).
  Notice, from (\ref{dotS}) and (\ref{dotO}), that symmetric initial conditions
  ($\bm{\Omega}|_{t=0}=0$) only evolve into symmetric dynamics (pure strain),
  while this is not true for the antisymmetric counterpart (pure rotation),
  because starting from $\bm{S}|_{t=0}=0$ a strain rate can develop.
  It is also worth mentioning that the symmetric pressure Hessian has no direct influence on vorticity:
  this is why, in (\ref{dotO}), the time scale $\tau$ only appears
  in the dissipative terms (which lack the order $\tau^1$), while no
  contribution from the pressure Hessian is present.
  On the other hand, the latter is the origin of the $O(\tau)$ term
  in the last line of (\ref{dotS}), which is not derived from viscosity
  (indeed it does not depend on $T$) but has been written in this way for the sake of simplicity.

  The simplification induced by the projection on $\bm{S}$ and $\bm{\Omega}$
  allows now to identify a linear term, multiplied by the term inside the curly parentheses.
  The sign of this term can then be investigated:
  we will later return to this issue in section \ref{sec:lin},
  in the framework of a simplified dynamics, obtained by discarding
  the $O(\tau^2)$ (or higher) terms in (\ref{expan}).

  \subsection{Unforced dynamics}

\begin{figure}
   \centering
   \includegraphics[scale=.65]{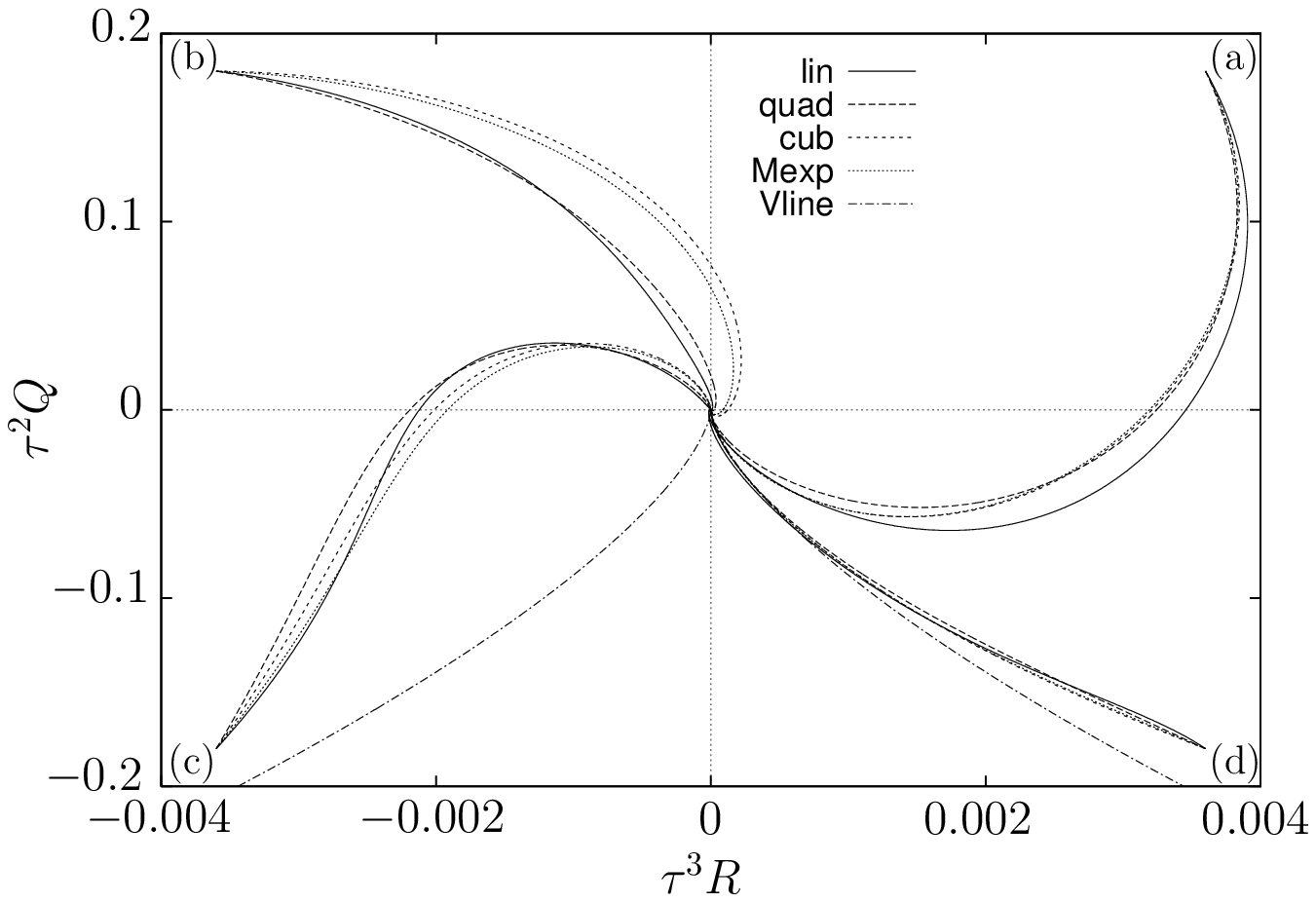}
   \caption{Evolution of $R$ and $Q$ (multiplied by the appropriate power of $\tau$)
    for the unforced dynamics using the RFD closure with $\tau=10^{-1}$,
    imposing the full matrix-exponential model or its linear, quadratic and cubic approximations.
    The four arbitrary initial conditions have been chosen as:
    (a) $A_{ij}|_{t=0}=3i-2j$, such that $Q|_{t=0}=18$ and $R|_{t=0}=36$;
    (b) $A_{ij}|_{t=0}=-3i+2j$, such that $Q|_{t=0}=18$ and $R|_{t=0}=-36$;
    (c) $A_{ij}|_{t=0}=3i-2j$ (except for $A_{31}|_{t=0}=-5$ and $A_{32}|_{t=0}=-7$) such that $Q|_{t=0}=-18$ and $R|_{t=0}=-36$;
    (d) $A_{ij}|_{t=0}=3i-2j$ (except for $A_{31}|_{t=0}=-5$ and $A_{32}|_{t=0}=-1$) such that $Q|_{t=0}=-18$ and $R|_{t=0}=36$
    (in all cases we have then redefined $A_{33}|_{t=0}\equiv-A_{11}|_{t=0}-A_{22}|_{t=0}$).}
   \label{unf-1}
  \end{figure}

\begin{figure}
   \centering
   \includegraphics[scale=.61]{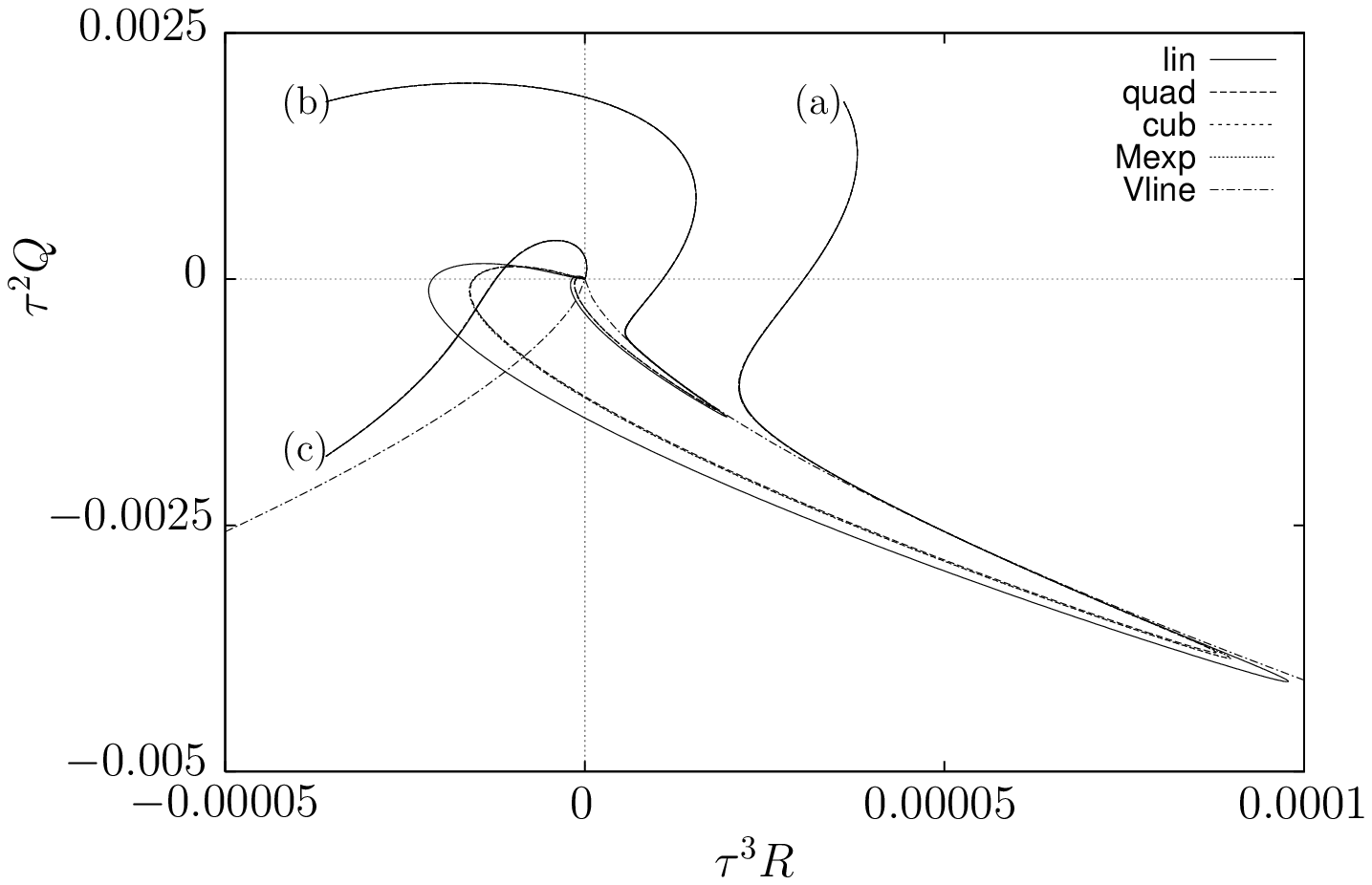}
   \caption{Same as in figure \ref{unf-1} using $\tau=10^{-2}$, but not including case (d), which has not been reproduced to avoid crowding in the figure.}
   \label{unf-2}
  \end{figure}

  Next, we wish to compare the numerical results of the full matrix-exponential approach (to be labeled as \textsf{Mexp})
  with the ones obtained by integrating (\ref{expan}),
  in which one can retain e.g.\ up to the first, second or third
  order in $\tau$ (labelled, respectively, as \textsf{lin}, \textsf{quad}, \textsf{cub}).\\
  In figures \ref{unf-1}--\ref{unf-2} we show the unforced evolution for the
  full matrix-exponential model and for its linear, quadratic and
  cubic approximations, starting from a few representative initial conditions in each of the quadrants of the $R$-$Q$ plane.
  Figure \ref{unf-1} shows the resulting time evolution for $\tau=10^{-1}$,
  for which all these unforced cases converge and decay towards the origin.
  For $\tau=10^{-2}$ (figure \ref{unf-2}) similar results can be obtained, although the weaker restitution term provided by the
  lower value of $\tau$ allows some excursions along the bottom-right Viellefosse tail before the decay towards the origin.
  These trends also hold at $\tau=10^{-3}$ (not shown), also starting from different quadrants,
  but then numerical instabilities occur if $\tau$ is further reduced,
  and these can only be avoided if the integration time step is reduced drastically.

  The important observation is that, in all of the previous cases,
  the linear, quadratic and cubic expansions are similar to one another,
  and provide a good approximation of the full matrix exponential.
  Therefore, in what follows, we will only focus on the comparison
  between the linearized model and the full, original one,
  not focusing on the $O(\tau^2)$ and $O(\tau^3)$ approximations.

  \subsection{Forced dynamics}

\begin{figure}
   \centering
   \includegraphics[scale=.6]{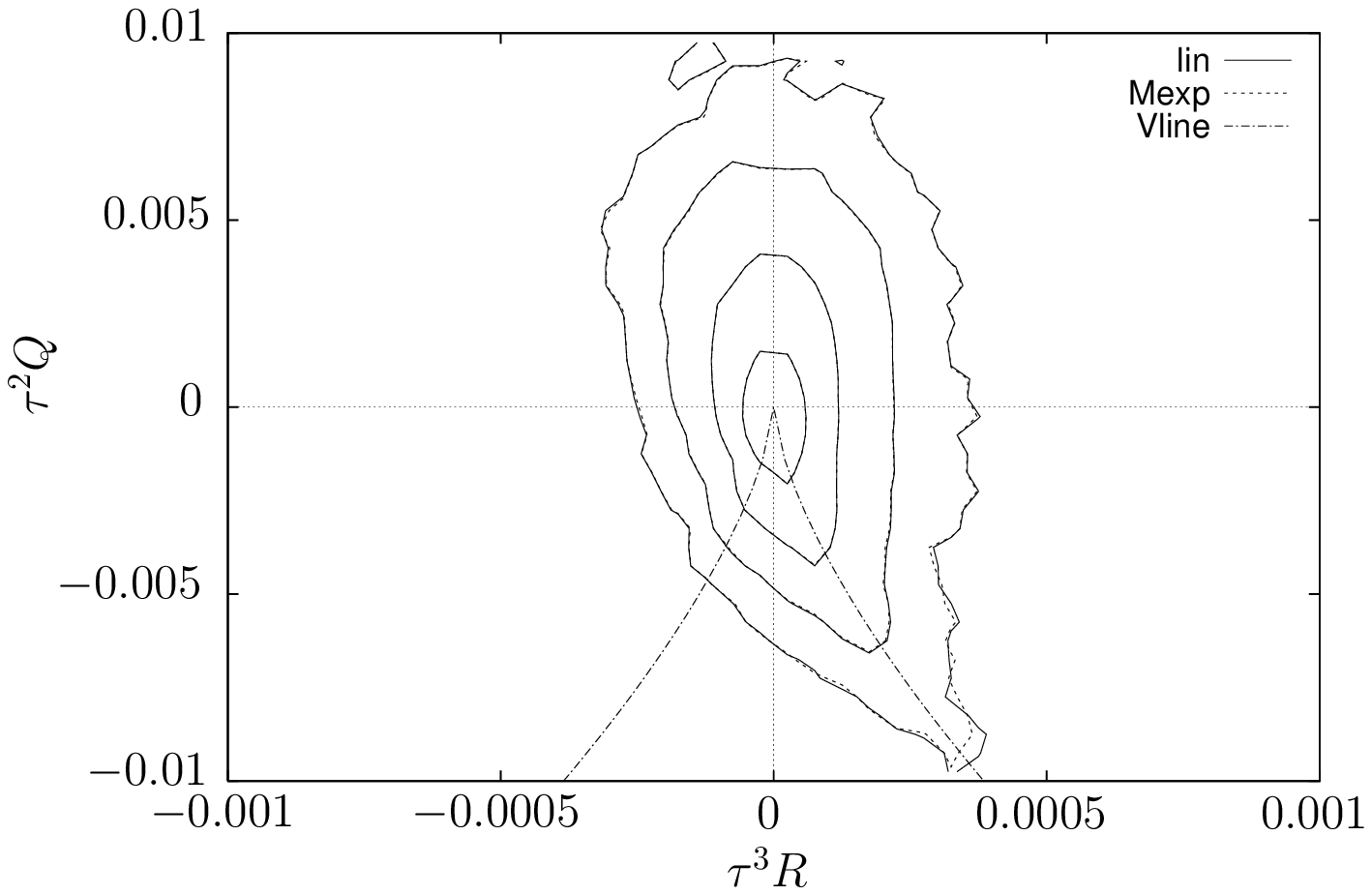}
   \caption{Same as in figure \ref{fig:m-1} with
    both the full matrix-exponential model and its linear approximation shown.
    Proceeding from the interior outwards, the isolines are for densities of $10$, $1$, $0.1$ and $0.01$ respectively.}
   \label{fig:-1}
  \end{figure}

  Let us now introduce again our additive forcing term,
  with a finite correlation time $\theta=\tau$
  (according to the framework described in section \ref{sec:num}
  and in appendix \ref{sec:app}),
  also on the RHS of (\ref{expan}) only including the first order in $\tau$.\\
  At $\tau=10^{-1}$, and $\mathcal{F}=1$, the linear approach is
  basically equivalent to the (already analyzed) matrix-exponential
  one in terms of resulting PDF's, as shown convincingly in figure \ref{fig:-1}.
  \\
  Reducing $\tau$, the simulations with the linear
  approximation based on the energy-dissipation criterion
  ($\mathcal{F}=4.5$ for $\tau=10^{-2}$ and
  $\mathcal{F}=14.35$ for $\tau=10^{-3}$)
  are again basically equivalent to the ones from the full matrix
  exponential shown in figures \ref{fig:m-2}--\ref{fig:m-3}.
  However, if one tries to reproduce the dimensional-scaling-based
  results of figures \ref{fig:m2}--\ref{fig:m3}
  ($\mathcal{F}=31.6$ for $\tau=10^{-2}$ and
  $\mathcal{F}=1000$ for $\tau=10^{-3}$),
  simulations with the linear approximation begin to suffer from numerical instabilities,
  regardless of how small the time step is chosen.
  It is therefore not possible to perform detailed numerical comparisons for significantly lower values of $\tau$.
  In order to proceed one step further using analytical tools, in the next section, the evolution equation
  for the complete set of relevant invariants is studied.

 \section{Analytical study of the invariants in the ``linearized'' equations} \label{sec:lin}

  Focusing on the ``linearized'' model, derived from (\ref{expan})
  by retaining only up to the first-order terms in $\tau$,
  allows us to proceed one step further analytically. Following \cite{MDV98},
  we derive a dynamical system for the five tensor invariants  $Q$, $R$, $Q_{\bm{S}}$, $R_{\bm{S}}$ (already introduced
  in section \ref{sec:exp}), and $V^2$, the latter defined from the strain-vorticity
  alignment $V=|\bm{S}\cdot\bm{\omega}|$, where
  $\omega_i\equiv-\epsilon_{ijk}\Omega_{jk}$ is the  vorticity vector.
  Notice that $Q_{\bm{S}}\le0\le V^2$ by definition, moreover
  it can easily be shown that $Q_{\bm{S}}\le-(27\,{R_{\bm{S}}}^2/4)^{1/3}$,
  so that the dynamics is confined below the zero-discriminant curve
  in the $R_{\bm{S}}$-$Q_{\bm{S}}$ plane \cite{MDV98}.
  It is worth mentioning that it is not necessary to investigate
  the antisymmetric counterparts of $Q_{\bm{S}}$ and $R_{\bm{S}}$,
  because by definition $Q_{\bm{\Omega}}=Q-Q_{\bm{S}}$
  and $R_{\bm{\Omega}}=0$ \cite{MDV98}.

  Starting from the equation for $A_{ij}$ (without forcing) and using appropriate contractions and the Cayley--Hamilton theorem,
  it is possible to obtain the following system of five equations:
  \begin{equation} \label{qsrsv}
   \left\{
   \begin{array}{rcl}
    \ud_tQ&=&\displaystyle-3R-2T^{-1}\left[1-\frac{4}{3}Q_{\bm{S}}T\tau\right]Q\\
    \ud_tQ_{\bm{S}}&=&\displaystyle-R-2R_{\bm{S}}-2T^{-1}\left[1-\frac{4}{3}QT\tau\right]Q_{\bm{S}}\\
    \ud_tR&=&\displaystyle\frac{2}{3}Q^2+\frac{8}{3}QR_{\bm{S}}\tau-3T^{-1}\left[1-\frac{4}{9}Q_{\bm{S}}T\tau\right]R\\
    \ud_tR_{\bm{S}}&=&\displaystyle\frac{2}{3}QQ_{\bm{S}}+\frac{1}{4}V^2-3T^{-1}\left[1-\frac{4}{3}Q_{\bm{S}}T\tau\right]R_{\bm{S}}\\
    \ud_tV^2&=&\displaystyle\frac{16}{3}Q(R-R_{\bm{S}})-4T^{-1}\left[1-\frac{2}{3}Q_{\bm{S}}T\tau\right]V^2\;.
   \end{array}
   \right.
  \end{equation}
  Notice that, in order to keep track of the various terms, we have not non-dimensionalized the equations and kept the term $T$.
  System (\ref{qsrsv}) shows linear and quadratic couplings among the different variables.
  What is most interesting is the fact that the linear ``damping'' terms,
  which have been factored at the end of each line for the sake of clarity,
  include terms at $O(\tau)$. It is of particular interest to study
  the sign of the square braces: when they (or at least one of them) becomes negative,
  divergences and blow-ups are likely to occur. However, in all equations except the one for $Q_{\bm{S}}$, these
  terms are positive definite (in fact larger than 1) since $Q_{\bm{S}}$ is never positive by definition. The possibility of
  negative damping is possible in the equation for $Q_{\bm{S}}$ itself, in which the damping term may become negative when
  $Q>3/(4T\tau)$, that is in highly rotating regions when $Q$ is large and positive.
  In fact, the rate of phase-space contraction ($-\nabla\cdot\bm{z}$,
  where $\bm{z}$ is the vector formed by the RHS of (\ref{qsrsv})) is
  $14T^{-1}-40Q_{\bm{S}}\tau/3-8Q_{\bm{\Omega}}\tau/3$, i.e.\ for
  overall growing solutions one would need
  $Q_{\bm{\Omega}}>5|Q_{\bm{S}}|+21/(4T\tau)$,
  very large rates of rotation.
  Clearly, however, along the excursions of the system along the Vieillefosse tail
  (the right-bottom part of the $R$-$Q$, where the RE model is known to diverge) when
  $Q$ is negative, the term provides damping and thus protects against divergencies along this direction.

  To obtain a system that generates stationary PDF's it is difficult to introduce random noise properly on the RHS of (\ref{qsrsv}).
  The difficulty stems from the inherent bounds and internal consistency conditions among the
  tensor invariants (e.g.\ the limits on $Q_{\bm{S}}$ and $V^2$ mentioned above). Simply adding random noise to the
  equation system \ref{qsrsv} quickly leads to $Q_{\bm{S}}>0$ or $V^2<0$, etc..
  Introducing a simple, additive random forcing on the RHS of the
  evolution equation for $\bm{A}$, $\ud_t\bm{A}=\ldots+\bm{F}$,
  leads to unclosed terms on the RHS of (\ref{qsrsv}). Therefore, the analytical study of the invariants is limited to the
  results presented above.

  Another set of interesting variables can be constructed by considering the tensor invariants of $\bm{A}$
  combined with a material line element $\bm{r}$ being convected by the flow. Specifically,
  following \cite{LM05,LM06}, one can introduce $\delta u\equiv\ell A_{ij}r_ir_j/r^2$ (which
  can have either sign) and $\delta v\equiv|\ell(\delta_{ij}-r_ir_j/r^2)A_{jk}r_k/r|$
  (the magnitude of the transverse velocity vector $\delta {v_i}$). These two variables represent the longitudinal and transverse
  velocity increments at a fixed scale $\ell$, i.e.\ between two fixed points with constant
  separation (rather than between two material points or fluid particles
  whose relative distance $r(t)$ would change with time). The derivation of a Lagrangian time evolution
  equation for $\delta u$ and $\delta v$ based on the equations for $\bm{A}$ and $\bm{r}$ leads to the
  ADV system introduced and studied in \cite{LM05,LM06}. In this prior work, it was shown that
  the neglect of viscous and much of the pressure Hessian term leads to unphysical behavior of the dynamics.
  Hence, it is of interest to consider the implications of the RFD closure and its expansions in the context of the ADV system.

  Repeating the relevant derivations \cite{LM05,LM06} but including the RFD closure
  expanded to first order in the equation for $\bm{A}$, one obtains:
  \[\left\{\begin{array}{ll}\ud_t\delta u=&\displaystyle-\frac{1}{\ell}\delta u^2+\frac{1}{\ell}\delta v^2-\frac{2}{3}\ell Q-\frac{1}{T}\left[1-\frac{4}{3}QT\tau\right]\delta u\\\ud_t\delta v=&\displaystyle-\frac{2}{\ell}\delta u\delta v-\frac{1}{T}\left[1-\frac{2}{3}QT\tau\right]\delta v\\&\ +\frac{2}{3}Q\tau A_{ij}r_i\delta v_j\delta v^{-1}\;.\end{array}\right.\]
  While the equation for longitudinal velocity increment is closed (if $Q$ is known), it is apparent that the
  equation for transverse velocity increment cannot be closed in terms of $\delta u$, $\delta v$ and $Q$.
  As shown in \cite{LM06}, parts of the invariant $Q$ may be expressed directly in terms of $\delta u$, namely
  $Q=-\delta u^2/\ell^2+Q^{-}$, where $Q^{-}$ is composed of terms that may not be expressed in terms of $\delta u$ or $\delta v$.
  The argument above is useful, since it shows that within the approximation $Q^{-}=0$ (as was done in \cite{LM06}),
  the added term from the linear expansion of the
  pressure Hessian model is positive damping since $(1-4QT\tau/3)=(1+4\delta u^2T\tau/3\ell^2)>0$, independent of $\delta u$.
  Hence, by setting $Q^{-}=0$ and neglecting the last term in the equation of the transverse velocity increment, the following
  system corresponds to the order $\tau^1$ RFD closure applied to the ADV system (in three dimensions):
  \begin{equation} \label{linADV}
   \left\{
   \begin{array}{ll}
    \ud_t\delta u=&\displaystyle-\frac{1}{3\ell}\delta u^2+\frac{1}{\ell}\delta v^2-\frac{1}{T}\left[1+\frac{4T\tau}{3\ell^2}\delta u^2\right]\delta u\\
    \ud_t\delta v=&\displaystyle-\frac{2}{\ell}\delta u\delta v-\frac{1}{T}\left[1+\frac{4T\tau}{3\ell^2}\delta u^2\right]\delta v\;.
   \end{array}
   \right.
  \end{equation}

 \section{Conclusions and perspectives} \label{sec:conc}

  In this work we have explored several consequences of the RFD closure as applied to modeling
  pressure Hessian and viscous terms for the Lagrangian dynamics of the velocity gradient tensor.
  Consistent with the observations of \cite{CM06,CM07}, the model is shown to predict unphysical dynamics when
  attempting to reach high Reynolds numbers (decreasing the time-scale parameter $\tau$). In particular, for
  $\tau<10^{-2}$, results quickly deteriorate. Analysis of the time-correlation structure of the solution allowed us
  to conclude that an inconsistency develops, in which the assumed correlation time scale ($\tau$) becomes much smaller than
  the actual correlation time scale predicted by the dynamics of the system. Hence, at decreasing $\tau$, the assumptions
  underlying the closure are not consistent with the simulated dynamics. How to ``break'' the observed longer-than-expected
  time correlations of the system is not clear and requires further study. Increasing the forcing strength was not a solution to the problem.
  In fact, a number of experiments were performed and in no case was it possible to obtain physically
  realistic predictions at high Reynolds number for $\tau<10^{-2}$. We must tentatively conclude that it may be
  impossible for a ``single-shell'' model in which only the dynamics at the smallest scales
  (largest velocity gradients) are computed dynamically to provide accurate predictions at arbitrarily high Reynolds numbers.
  As $\mathrm{Re}$ grows, it may be impossible to describe the dynamics with only 8 independent degrees of freedom, even
  though it is already remarkable that at moderate $\mathrm{Re}$ such a system is able to reproduce
  many features of turbulence \cite{CM06,CM07}. This was the motivation for proposing a shell model \cite{BCMT07} that included
  additional degrees of freedom.

  With the aim at improving our understanding of the dynamical effects of the recent fluid deformation closure,
  an expansion into powers of the small parameter (correlation time scale) was performed. It showed to be a good
  approximation, at least for $\tau>10^{-2}$. It allowed to show that, to first order, the RFD model for the pressure Hessian is
  a dissipative ``damping'' term for most of the tensor invariant's time evolution, except for the second invariant of the strain-rate
  tensor in regions of high rotation. That is also the region in which direct comparisons with DNS in \cite{CMBT08} showed significant
  inaccuracies of the closure. The effect of the linearized RFD closure for the pressure Hessian was shown to lead to a
  positive (dissipative) damping in the ADV equations (\ref{linADV}) for longitudinal and transverse velocity increments.
  For future efforts, it would be of interest to study the properties of this simple system of equations.
  In particular, one may want to develop stochastic forcing for the system in order to test whether it
  may yield realistic stationary statistics for longitudinal and transverse velocity increments in turbulence.

 \section*{Acknowledgements}

  We thank Laurent Chevillard for useful discussions and suggestions. M.M.A.\ is supported by
  postdoctoral Fellowship from the Keck Foundation and C.M.\ by the National Science Foundation.
  The authors are delighted to dedicate this article to Prof.\
  K.R.\ Sreenivasan following the Symposium on Fluid Science and
  Turbulence held on occasion of his $60^{\mathrm{th}}$ birthday.

 \appendix

 \section{Tensorial random forcing term} \label{sec:app}

  In this appendix, we describe in more detail the forcing $\bm{F}$ used in sections \ref{sec:num} and \ref{sec:exp}.
  $\bm{F}$ is a random, traceless, tensorial noise that is restricted to comply with various tensorial symmetries to be
  consistent with those of the velocity gradient tensor.

  At first, we generate random deviates with uniform probability distributions,
  and turn them into normal deviates by means of a simple transformation,
  specifying its Jacobian \cite{PFTV93}. This is done for each of the
  9 tensor components. Then, in order to obtain the white noise forcing used
  in some preliminary tests, and then as a comparison in figures
  \ref{fig:s-1}--\ref{fig:comp}), we adopt the procedure explained
  in appendix A of \cite{CMBT08}. This procedure allows us to obtain the correct properties
  for the trace of the tensor, and for the variances of the longitudinal
  and transverse components (the latter are the double of the former).

  Next, we describe how to obtain the time-correlated noise.
  A time-uncorrelated tensorial noise ($\bm{W}$) is used to obtain a
  finite-correlated ($\bm{F}$) noise using an Ornstein-Uhlenbeck process \cite{G85}:
  \begin{equation} \label{oup}
   \ud\bm{F}=-\theta^{-1}\bm{F}\,\ud t+\mathcal{F}\theta^{1/2}\,\ud\bm{W}\;,
  \end{equation}
  where $\theta$ and $\mathcal{F}$ are the desired correlation time
  and typical amplitude, respectively, for $\bm{F}$.
  The correlation time will be set equal to $\tau$
  (tests with $\theta=\tau/2$, $=\tau/5$ and $=\tau/10$
  were also presented, together with the white-noise case $\theta=0$).
  As explained in section \ref{sec:num}, the appropriate value of
  $\mathcal{F}$ to be prescribed
  is found empirically through figure \ref{fig:points} (and is corrected by introducing a multiplicative factor in
  subsection \ref{sec:timecorr} for various $\theta$ values), and is such as to satisfy an energy-dissipation-based criterion.
  Note that the root-mean-square values of the transverse
  components are equal to $\mathcal{F}$, while a further factor $1/\sqrt{2}$ is present
  for the longitudinal ones (e.g., $F_{11}^{\mathrm{rms}}\simeq F_{12}^{\mathrm{rms}}/\sqrt{2}$).
  Also, a single value of $\mathcal{F}$ must be used in (\ref{oup}) for every component,
  because the correct ratio between the longitudinal and transverse amplitudes
  is guaranteed by the already-mentioned corresponding ratio for the white-noise components.

\end{document}